\documentclass[12pt,preprint]{aastex}
\usepackage{epsfig}
\begin{document}
\voffset-1cm
\newcommand{\gsim}{\hbox{\rlap{$^>$}$_\sim$}}
\newcommand{\lsim}{\hbox{\rlap{$^<$}$_\sim$}}

\title{Short Hard Gamma Ray Bursts And Their Afterglows}

\author{Shlomo Dado\altaffilmark{1} and Arnon Dar\altaffilmark{2}}

\altaffiltext{1}{dado@phep3.technion.ac.il\\
Physics Department and Space Research Institute, Technion, Haifa 32000,
Israel}
\altaffiltext{2}{arnon@physics.technion.ac.il,
dar@cern.ch\\
Physics Department and Space Research Institute, Technion, Haifa 32000,
Israel\\ Theory Unit, CERN, 1211 Geneva 23, Switzerland }

\begin{abstract}

Long duration gamma ray bursts (GRBs) and X-ray flashes (XRFs) are 
produced by highly- relativistic jets ejected in core-collapse supernova 
explosions. The origin of short hard gamma-ray bursts (SHBs) has not been 
established. They may be produced by highly relativistic jets ejected in 
various processes: mergers of compact stellar objects; large-mass 
accretion episodes onto compact stars in close binaries or onto 
intermediate-mass black holes in dense stellar regions; phase transition 
of compact stars. Natural environments of such events are the dense cores 
of globular clusters, superstar clusters and young supernova remnants. We 
have used the cannonball model of GRBs to analyze all Swift SHBs with a 
well-sampled X-ray afterglow. We show that their prompt gamma-ray emission 
can be explained by inverse Compton scattering (ICS) of the progenitor's 
glory light, and their extended soft emission component by ICS of high 
density light or synchrotron radiation (SR) in a high density interstellar 
medium within the cluster. The mechanism generating the afterglow is 
synchrotron radiation outside the cluster. No associated supernova could 
be detected in the low luminosity nearby GRBs 060614 and 060505. We 
interpret them as SHBs seen relatively far off axis.

\end{abstract}
\maketitle

\section{Introduction}

Gamma ray bursts (GRBs) have been traditionally classified into short hard 
bursts (SHBs) and long soft bursts. Their distribution as a function of 
duration is bimodal with a minimum around two seconds (Mazets et al.~1981; 
Norris et al.~1984, Kouveliotou et al.~1993). The $\gamma$-rays of short 
bursts are typically harder than those of long bursts, thus their acronym, 
SHBs, but their fluence is much smaller. The short hard bursts consist of 
a single or a complex of peaks with a total duration typically less than 2 
seconds and peak widths which range between a few ms and a fraction of a 
second. The lag-time between their emission in soft and hard energy band 
is much smaller than that in long GRBs and X-ray flashes (XRFs).
We shall refer to the ensemble of long GRBs and XRFs as LGRBs.

A true breakthrough in the observations of SHBs came with the launch of 
Swift in November 2004. Since its launch Swift has detected over 30 SHBs 
with 
X-ray and UVO afterglows, and triggered follow-up ground-based 
observations that for most SHBs succeeded in identifying their 
host galaxy and measuring its  
redshift. Swift has also discovered that, in more than 25\% of these 
cases, perhaps in all, the SHB was followed by an extended soft emission 
component (ESEC), a spectrally softer component lasting tens of seconds 
(for recent reviews see Nakar~2007, Kann et al.~2008). 
The fluence of SHBs is typically much 
smaller than that of GRBs. The fluence of their ESEC is usually comparable 
to or smaller than that of the SHB. The ESECs end with a fast-decay phase, 
followed by an afterglow with a canonical time- dependence, similar to the 
one observed in a large fraction of LGRBs (Nousek et al.~2006). LGRBs 
are located near the center of their hosts, which are all late-type 
galaxies (Covino et al.~2006). Contrariwise, SHBs have an heterogeneous 
population of hosts, and take place both in elliptical galaxies (e.g., 
SHBs 
050709 and 050724, Gehrels et al.~2005, Berger et al.~2005) and spiral 
galaxies (e.g., SHB 051221A, Soderberg et al.~2006c). Generally, the SHBs 
with an ESEC are centrally located, while those with no detectable ESEC 
are found at very large distances from their host's center (Troja et 
al.~2008). There is also a clear trend of their afterglow to 
be fainter for larger offsets (Kann et al.~2008).

Observations indicate that LGRBs are mainly produced by core-collapse 
supernovae (SNe) of type Ib/c. A GRB-SN association was discussed long ago 
(Colgate 1968, Dar et al.~1992) but was 
dismissed by Woosley~(1993) who suggested that only `failed supernovae' 
produce GRBs.  Following the discovery by Galama et al.~(1998) of 
SN1998bw in the error box of GRB980425, 
Wang \& Wheeler~(1998), Dar \& Plaga~(1999) and Dar \& De 
R\'ujula~(2000) suggested that perhaps most core collapse SNe produce 
LGRBs. This was later advocated as an observed fact, based on a 
comprehensive analysis of optical afterglows within the cannonball (CB) 
model of LGRBs (Dado, Dar \& De R\'ujula~2002,2003 hereafter 
DDD2002.DDD2003, Dar \& De R\'ujula~2004 hereafter DD2004,  and references 
therein) and on empirical grounds by Zeh, Klose \& Hartmann~(2004). 
General acceptance of the GRB-SN association waited until  the 
spectroscopic discovery of an SN2003dh coincident with GRB030329 (Hjorth 
et al.~2003, Stanek et al.~2003) and b additional discoveries of 
spectroscopically proven GRB-SN associations such as GRB030213/SN2003lw 
(Malesani et al.~2003), GRB021211/SN2002lt (Della Valle et al.~2003), 
XRF060218/SN2006aj (Campana et al.~2006a, 
Pian et al.~2006, Mazzali et al.~2006) and XRF080109/SN2008D (Malesani et 
al.~2008, Modjaz et al.~2008). 

In contrast, SHBs do not seem to be 
associated with any known type of supernova (e.g., Hjorth et 
al.~2005b). 
The identity of their progenitors is not established, except that giant 
flares from soft gamma ray repeaters (SGRs) that look like SHBs, may 
account for most of the observed SHBs in nearby  (redshift $z\lsim 
0.01$) galaxies 
(Dar~2005, Hurley et al.~2005).  The extremely bright SHBs 051103 and 
070201 observed by 
Konus-Wind could have been such events in the nearby 
galaxies M81/M82 and M31 (Ofek et al.~2006, Frederiks et al.~2007). But 
hyperflares from SGRs cannot account for the bulk of the much more distant 
ones, $z \!\gg\! 0.01$, unless they are highly collimated or 
much more energetic but less frequent.

Although the observational data on cosmological SHBs and their afterglows 
did not pin down their progenitors, they do provide useful clues to their 
progenitors, production mechanism, and production sites. Their extremely 
short durations and large equivalent isotropic gamma ray energies, and the 
absence of self absorption features in their $\gamma$-ray spectra suggest
that they are produced by highly relativistic jets ejected in violent 
processes involving compact stars (Shaviv \& Dar 1995a) such as:

\begin{itemize}
\item{}
Neutron stars merger and neutron star - black hole merger in compact 
binaries (Blinnikov et al.~1984, Paczynski~1986, Goodman, Dar \&
Nussinov~1987, Eichler et al.~1989).

\item{} 
Collapse of compact stars (neutron stars, hyper stars, quark stars) to a 
more com-
pact star due to mass accretion, and/or loss of angular momentum and/or 
cooling by
radiation (Dar~1998a,1999, Dar \& De R\'ujula~2000, Dar~2006).
 
\item{}
Phase transition in compact stars, such as neutron-stars, hyper-stars and 
quark stars (Dar~1999, Dar \& De R\'ujula~2000, Dar~2006).
 
\item{}
Accretion episodes in microblazars (Dar~1998b,1999) 
and on intermediate mass black holes in dense stellar regions.

\end{itemize}

The natural environment of these progenitors are super dense stellar 
regions such as collapsed cores of globular clusters (GCs) and young 
centrally-condensed super-star cluster (SSC) (Shaviv \& Dar 1995b) whose 
stellar densities and total luminosities seem to be correlated with their 
distance from the center of their host galaxy. If the ESEC is produced by 
ICS of ambient light in globular clusters with a low density interstellar 
medium (ISM) or by synchrotron radiation (SR) 
in super-star clusters with a very large ISM density, it may explain the 
observations that SHBs with an ESEC are centrally located, while those 
with no detectable ESEC are found at very large distances from their 
host's center (Troja et al.~2008).

The spectra and pulse-shapes of SHBs measured by the $\gamma$-ray 
telescopes on board satellites, except for being harder and shorter, are 
very similar to those measured by the same satellites in LGRBs 
(Nakar~2006, Kann et al.~2008). In particular, the X-ray light curves of 
SHBs, which were well sampled with the Swift Burst Alert Telescope (BAT) 
and X-ray Telescope (XRT) show the same canonical behaviour seen in many 
LGRBs (Nousek et al.~2006, O'Brien et al.~2006): An early rapid temporal 
decay of the prompt emission during which the spectrum softens at a very 
fast rate. This rapid decay phase ends within a few hundred seconds when 
it is taken over by a plateau with a much harder power-law spectrum, 
typically lasting thousands to tens of thousands of seconds . Within a 
time of order of one day, the plateau steepens into a power-law decay, 
which 
lasts until the X-ray AG becomes too dim to be detected. Often, X-ray 
flares, not coinciding with any detectable $\gamma$-ray activity, 
are 
superimposed on the X-ray light curve during the fast-decaying phase and 
even later. These similarities with long GRBs suggest that the initial 
burst and the afterglow in both SHBs and LGRBs are produced by the same 
radiation mechanisms: inverse Compton scattering (ICS) of glory light and 
synchrotron radiation (SR) from decelerating CBs.

The CB model has been successful in predicting/accommodating the observed 
properties of the prompt radiation of LGRBs (Dar \& De 
R\'ujula~2004 - hereafter DD2004) and their afterglows (Dado, Dar \& De 
R\'ujula 2002, 2003, 2007b, 2008a,b - hereafter DDD2002, DDD2003, DDD2007b,
DDD2008a,b,c). Because of the above similarities between LGRBs 
and SHBs, the CB model may also reproduce 
correctly the main observed properties of SHBs, after only simple 
adjustments. Indeed, in this paper we show that using only general 
properties of the putative SHB progenitors and their natural environments, 
the CB model with the same two basic radiation mechanisms, ICS of ambient 
light, which dominates the prompt emission, and synchrotron radiation 
which 
dominates the afterglow emission, provides also a simple and successful 
description of the observed properties of SHBs and their afterglows. 
Moreover, a few nearby LGRBs without a detectable supernova, such as GRB 
060614, which were interpreted as belonging to a new class of LGRBs (e.g., 
Gal-Yam et al.~2006b, Gehrels et al.~2006, Della Valle et al.~2006) can be 
interpreted as SHBs seen far off-axis.

\section{Summary of the underlying assumptions}

\subsection{The progenitors of GRBs and their environments}

We have no novel suggestions regarding either the putative progenitors of 
cosmological SHBs or the ejection mechanism and the properties of their 
highly relativistic jets. As listed in the introduction, most 
conventionally, the progenitors may be the merger of a neutron star 
with another one or with a black hole, or large accretion episodes in 
microblazars (microquasars with their jets pointing accurately to us) or 
intermediate mass black holes. Accretion, cooling or angular momentum loss 
may result in abrupt transitions of neutron stars to more compact objects 
(strange stars, quark stars or stellar black holes). All these phenomena 
are natural candidates for SHB progenitors. The above putative progenitors 
meet the energy-budget requirements, even if the radiation of SHBs is not 
very highly collimated. The estimated merger rate of neutron stars is 
sufficient only if SHBs are not narrowly collimated. If they are, as in 
the CB model, some of the other mechanisms we cited must also be 
operative. Indeed, low-luminosity LGRBs detected by INTEGRAL led to the 
conclusion (Foley et al.~2008) that the rate of LGRBs with inferred low 
luminosity is $\gsim 25\%$ of that of Type Ib/c supernovae\footnote{Such 
a fraction is larger by a factor $\sim 10^{2}\!-\!10^{3}$ than that of 
previous estimates (e.g., Gal-Yam et al.~2006a and Soderberg et 
al.~2006b), 
which relied on the validity of uncertain fireball model relations.}. 
Since BATSE observed an SHB rate nearly 1/3 the rate of LGRBs, the rate of 
SHBs implied by the INTEGRAL observations must also be comparable to the 
rate of SNe of type Ib/c. Such a rate is a considerable fraction of the 
birth rate of neutron stars.  It is much larger than the merger rate of 
neutron stars in close binaries. It favours other alternative sources of 
SHBs such as a phase transition in neutron stars or accretion episodes on 
stellar mass and intermediate mass, black holes.

The putative progenitors of SHBs are mostly located in the cores of 
globular clusters and the central region of superstar clusters. 
We shall assume that SHBs are produced in such an environmemt.

\subsection{The Radiation mechanisms}

We shall assume that similar to LGRBs, the two dominant radiation 
mechanism in SHBs are ICS and SR. The ICS of glory light produced by a 
stellar companion or by an accretion disk, by electrons comoving with the 
CBS, dominates the prompt emission, while SR from the decelerating CBs in 
the ISM or the intergalactic medium (IGM) dominates the afterglow 
emissions.

We shall also assume that ICS of ambient light in globular clusters of 
low density ISM, or SR from the decelerating CBs in the high density ISM 
of superstar clusters produces the ESEC.

\subsection{The main parameters}

The launching of highly relativistic and narrowly collimated
jets in mass accretion episodes, merger or phase transitions 
is not sufficiently understood to predict the chaotic timing of CB  
emissions, nor the number, $N_{CB}$, of dominant CBs in a particular jet, 
nor the baryon number, $N_B$, nor the initial Lorentz factor, $\gamma_0$, 
of observable CBs. But once the typical values of these parameters are 
inferred from the data, all properties of LGRBs and SHBs can be understood 
in simple terms. This also requires a number of items of 
information independent of the model itself. They are the angle 
between a CB jet and the observer, the properties of the ambient light a 
CB encounters in its flight, and of the density of the interstellar medium 
(ISM) or intergalactic medium (IGM)
which CBs pierce through. Naturally, the cosmological redshift, z, 
and the absorption or attenuation of light of various frequencies along 
the line of sight play their usual roles, with the exception, in 
comparison with other models, that the line of sight to a CB moves 
significantly (and hyper-luminally) during the observation time.
The standard cosmological model with $\Omega\!=\! 1$, $\Omega_M\!=\!0.27$,
$\Omega_\Lambda\!=\! 0.73$ and $H_0\!=\!71\, {\rm km\, s^{-1},\, 
Mpc^{-1}}$ is assumed throughout this paper. 

\subsection{The opening angle of a jet of CBs and the typical observers' 
angle}

We hypothesize that a CB initially expands in its rest system at a speed 
comparable to the speed of sound in a 
relativistic plasma, $v\!\equiv \! \beta_s\, c/\sqrt{3}$ with 
$\beta_s\lsim 1$. A CB's initial radius 
must be of the order of the size of the progenitor compact object or its 
accretion disk, and rapidly becomes negligible as the CB expands. Thus, a 
CB traces a cone in space of initial opening angle 
$\theta_{CB}\!=\!\beta_s/\sqrt{3}\, \gamma_0.$ The characteristic opening 
angle of 
the emitted radiation is $1/\gamma_0$, much larger than $\theta_{CB}$, so 
that 
the jet opening is not a concept that plays a very major role. In the AG 
phase, the effect of the deceleration of CBs in the ISM becomes important, 
and the radiation is spread over an angle $1/\gamma(t)$. 

In the CB model, the Doppler factor, $\delta(t)$, relating times, energies 
and fluxes in a CB's rest system to those in the observer's system plays a 
major role in the understanding GRBs. Its form in terms of $\theta$ and 
the time-dependent Lorentz factor, $\gamma(t)$, of a CB, is: 
\begin{equation} 
\delta(t)={1\over \gamma(t)\, (1-\beta(t)\, 
\cos\theta)}\approx {2\, \gamma(t)\over 1+[\gamma(t)\,\theta]^2}\; , 
\label{delta} 
\end{equation} 
where the approximation is excellent for 
$\gamma\gg 1$ and $\theta\ll 1$. Doppler boosting 
and relativistic beaming enhances
the observed energy flux from a CB  
by a factor $\delta^3$,
making the observations of GRBs increasingly improbable to 
observe GRBs at angles $\theta>1/\gamma$. The angular phase space being 
$d\Omega\propto \theta\, d\theta$, the most probable angle of
observation is $\!\approx\!1/\gamma$.

\section{The prompt emission}

\subsection{The spectrum of ICS pulses}

We contend that the prompt $\gamma$ and X-rays of an SHB's pulse are made 
similar to those in LGRBs, i.e., by ICS of glory light by the electrons 
contained within a CB. For LGRBs the glory light is the 
early 
flash of the SN explosion, scattered away from the radial direction by the 
nearby circumburst material ejected by pre-SN ejecta. For the prompt 
signal of SHBs, the glory light may be the light emitted by a companion 
star or 
a transient accretion disk and scattered by their winds.
During the initial phase of $\gamma$-ray 
emission in a SHB, the Lorentz factor $\gamma$ of a CB stays put at its 
initial value $\gamma_0$, for the deceleration induced by the collisions 
with the ISM has not yet had a significant effect. The electrons comoving 
with the CB scatter the photons of the glory around the progenitor, which 
has a thin thermal-bremsstrahlung spectrum,
\begin{equation}
\epsilon\, {dn_\gamma \over d\epsilon} \sim \left({\epsilon \over
\epsilon_g }\right)^{-\beta_g}\, e^{-\epsilon/\epsilon_g},
\label{thinbrem}
\end{equation}
with a spectral index $\beta_g\!\sim\!0$  (photon spectral index 
$\Gamma=\beta_g \!+\!1\!\sim\!1$)
and a typical (pseudo)-temperature,
$ \epsilon_g \!\sim\!2$ eV,  for a glory produced by stellar light.
The observed energy of a glory photon, which suffered an inverse
Compton scattering by an electron comoving with a
CB at redshift $z$, is then given by,
\begin{equation}
E={\gamma_0\, \delta_0\, \epsilon \, \over (1+z)}\, (1+\cos\theta_{in}),
\label{ICSboost}
\end{equation}
where $\theta_{in}$ is the angle of incidence of the initial
photon onto the CB, in the SN rest system.

\noindent
The density of the glory photons seen by a CB is given roughly by,
$n_g(r)\sim n_g(0)\,r_g^2/(r^2+r_g^2)$. At an early time,
$r\approx c\, \gamma_0\, \delta_0\, t/(1+z).$ Consequently,
$n_g(t)\sim n_g(0)\, \Delta t^2  /(t^2+ \Delta t^2)$,
where $\Delta t =(1+z)\, r_g/ c\, \gamma_0\, \delta_0$.
The factor  $\!1\!+\cos\theta_{in} $  in Eq.~(\ref{ICSboost})
must roughly average zero
near the center of the glory where the photon distribution
is semi isotropic and  tends to $1/r^2\! \sim\! 1/t^2 $ as the CB moves 
away.
This behaviour is described approximately by
$\langle 1\!+\!\cos\theta_{in}\rangle \approx 1\!-\!t/\sqrt{t^2+ \Delta 
t^2}$
and then the predicted time-dependent spectrum of the SHB pulse, produced
by ICS of the glory photons and observed at an angle $\theta$, 
is given by (DD2004):
\begin{equation}
E\, {dN_\gamma\over dE} \sim \left({E\over E_p(t}\right)^{-\beta_g}\,
 e^{-E/E_p(t)}+ b\,(1-e^{-E/E_p(t)})\, \left({E \over 
E_p(t)}\right)^{-p/2}\,,
\label{SHBspec}
\end{equation}
where,
\begin{eqnarray}
E_p(t)&\simeq& E_p(0)\, \left[1- {t\over \sqrt{t^2 + \Delta 
t^2}}\right]\,,
\nonumber\\
E_p(0)&=&{\gamma_0\, \delta_0 \over 1+z}\, \epsilon_g.
\label{PeakE}
\end{eqnarray}
The first term, with $\beta_g\!\sim\! 0$, is the result of Compton
scattering by the bulk of the CB's electrons, which are comoving with it.
The second term in Eq.~(\ref{SHBspec}) is induced by
a very small fraction of
`knocked on' and Fermi accelerated electrons, whose initial spectrum
(before Compton and synchrotron cooling) is $dN_e/dE\propto E^{-p}$,
with $p\approx 2.2$.
For $b\!=\!O(1)$, the predicted spectrum, Eq.~\ref{SHBspec}, bears a 
striking resemblance to the empirical Band
function (Band et al.~1993) traditionally used to model the energy spectra
of GRBs. GRBs which are well fit by the Band function are also well fit 
by Eq.~(\ref{SHBspec}), but GRBs where the spectral measurements 
extended over a much wider energy range than that of BATSE and Swift/BAT,
were better fit by Eq.~\ref{SHBspec} (e.g., Wigger et al.~2008).

For many Swift SHBs the spectral observations
do not extend to energies larger than $E_p(0)$,  or the value of $b$
in Eq.~(\ref{SHBspec}) is relatively small because of the low-density 
environment of the SHB,  
so that the first term of the equation, 
\begin{equation}
E\, {dN_\gamma\over dE} \sim \left({E\over E_p(t)}\right)^{-\beta_g}\,
 e^{-E/E_p(t)}\, .
\label{SHBspect}
\end{equation}
with $\beta_g\simeq 0$ ($\Gamma\!=\!1\!-\!\beta_g\simeq 1$) provides 
a very good 
approximation, and will be used 
in this paper. This term coincides with the `cut-off
power-law' spectrum which has been used to model the 
time integrated spectra of many LGRBs and SHBs. 
The resulting values of $\Gamma$ that are reported in Table I
for time integrated spectra, are all consistent within the observational 
uncertainties
with the ICS expected value, $\Gamma\!=\!1$.

\subsection{The light curve of the prompt ICS pulses}

After its launch, the fast expanding CB 
propagates in the progenitor's glory and wind on
its way  to the ISM. Its cross section increases, while both its
density and opacity, as well as  those of the wind, decrease.
The pressure of the incoming particles and 
radiation on its front surface  may distort its shape to 
look more like a disk.  Let $t$ 
be the time after launch of a CB in the rest frame of the progenitor.
Approximating the CB geometry by a cylindrical slab of radius $R$
and neglecting the  spread in arrival times of scattered glory
photons in the CB that entered it simultaneously,
the  rate of scattered glory photons is given by,
\begin{equation}
{dN_\gamma\over dt}  = \pi\, R^2(t)\, c\, n_g(t)\, 
       \int dt'\,c\, \sigma_{_T}\,n(t')\,
           e^{-\int_t^{t'} \sigma_{_T}\, n(t")\, c\, dt"}.
\label{Nslab}
\end{equation}
The  integration  yields
\begin{equation}
{dN_\gamma\over dt}  = c\, n_g(t)\,\pi\, R^2(t)\, [1-e^{-\sigma_T\, N(t)}],
\label{dNdtslab}
\end{equation}
where $N(t)$ is the effective column density of the expanding CB
encountered
by a photon which begins crossing it at a time $t$.

If $t$ denotes the time of arrival of photons seen by a distant 
observer, it is  related to the time $t_r$ in the progenitor's
rest frame  via $dt\!=\!(1+z)\, dt_r/\gamma_0\, \delta_0$. 
The relativistic motion of the CB that boosts the energy of
the ICS photons, also collimates them 
by a factor $dcos\theta/dcos\theta_r\!=\!\delta_0^2$. 
Thus, the pulse produced by a CB at redshift $z$ and luminosity 
distance $d_L$ is described approximately by,
\begin{equation}
E\,{d^2N_\gamma \over dt\, dE}\approx   {c\, n_g(t)\, (1+z)\,\gamma_0\,
     \delta_0^3\over 4\,\pi\ d_L^2}\,e^{-\tau_{_W}(t)}\,
 \pi\, R^2(t)\, 
 [1-e^{-4\,R_{tr}^2/3\,R^2(t)}]\, E\,{dN_\gamma\over dE}\,\, ,
\label{ICSPulse0}
\end{equation}
where $\tau_{_W}= \sigma_{_W} \int_r\, n(r)\,dr\sim a/t$
is the radiative opacity of the wind,
and $E\, dN_\gamma/dE$ is given by Eq.~(\ref{SHBspec}).
A wind's density-profile, $n(r)= n_0\, r_0^2/r^2$,
yields, $a(E)=\sigma_{_W}\, n_0\, r_0^2\, (1+z)/c\,\gamma_0\, \delta_0$.
At sufficiently high energies the wind's opacity is mainly due to
Compton scattering, i.e., $\sigma_{_W}\approx \sigma_{_T}$,
where $\sigma_{_T}$ is the Thomson cross section.
At X-ray energies it is mainly due to bound-free and bound-bound 
transitions,
and at energies below the threshold for atomic photo excitation
it is dominated by free-free transitions and dust scattering.

The initial rapid expansion of a CB
slows down as it propagates
through the wind and scatters its particles
(DDD2002, DD2004).  This expansion can be described  roughly by,
$R^2\approx R_{cb}^2\, t^2/(t^2\!+\!t_{exp}^2)$, where $R_{cb}$
is the asymptotic radius of the CB. Since $t_{tr} \ll t_{exp}$
(DD2004), Eq.~(\ref{ICSPulse0}) can be well approximated by,
\begin{equation}
E\,{d^2N_\gamma \over dt\, dE} \propto {\Delta t^2\, t^2\, 
e^{-a/t}\over
    (t^2+\Delta t^2)\, (t^2+t_{tr}^2)}\, E\,{dN_\gamma\over dE}\, , 
\label{ICSPulse1} 
\end{equation} 
For nearly transparent winds ($a \rightarrow 0$) and
$t_{tr}\approx \Delta t $,  Eq.~(\ref{ICSPulse1}) has an
approximate shape, 
\begin{equation}
 E\,{d^2N_\gamma \over dt\, dE} \propto {\Delta t^2\,t^2 \over 
(t^2+\Delta t^2)^2}\,E\,{dN_\gamma\over dE}\, .
\label{ICSPulse}
\end{equation}
Except for very early times, this shape is almost 
identical to that of the `Master' formula  advocated in (DD2004) for 
the prompt ICS pulses of GRBs and XRFs, which was shown to well 
describe  the prompt emission pulses of LGRBs (e.g., DDD2007b, 
DDD2008a,b),
including the rapid decay of the prompt emission  
and its a fast spectral softening (DDD2008a).

The observed light curve (energy flux density) of
a multi-pulse SHB dominated by ICS is given by,
\begin{equation}
F_{E} \approx
\Sigma_i\, A_i\Theta[t-t_i]\,  {\Delta t_i^2\,(t-t_i)^2 \over
((t-t_i)^2+\Delta t_i^2)^2}\, e^{-E/E_{p,i}(t-t_i)}\, .
\label{ICSlc}
\end{equation}
where the index `i' denotes the i-th ICS pulse produced by a 
CB launched at an observer time $t\!=\!t_i$, with a light curve
given approximately by Eq.~(\ref{ICSPulse}), and where,
\begin{equation}
A_i \approx {c\, n_g(0)\, \pi\, R_{CB}^2\,\gamma_0 \delta_0^3\, (1+z)
\over 4\, \pi\, D_L^2}\, .
\label{Ai}
\end{equation}
Thus, in the CB model each ICS pulse 
is effectively described by four parameters, $t_i,\, A_i,
\Delta t_i$  and $E_{p,i}(0)$,
which are best fit to reproduce its observed light curve.

\subsection{Effective spectral index and hardness ratio}

The effective spectral index of unabsorbed ICS pulses
in Eq.~(\ref{SHBspect}) is given by (DDD2008a),
\begin{equation}
\Gamma_i(E,t-t_i) =-E\; {d\,{\rm log} F_E\over
                           dE}=
                1+ \beta_g+{E\over E_{p,i}(t-t_i)}\,.
\label{Gamm}
\end{equation}
The hardness, defined as the ratio between the number of events 
in two energy bands, is generally reported, uncorrected for absorption,
for the Swift
25-150 keV  and  15-25 keV bands. It
can be  approximated by (DDD2008a),
\begin{equation}
{\rm HR}_i(t)=B_i\, e^{-\Delta E/ E_{p,i}(t-t_i)}\,
\label{HR}
\end{equation}
where $\Delta E$ is an effective interval between the bands.

\subsection{The peak energy of a GRBs pulse}

For $b\!\sim\! 0$ and $\beta_g\!\sim \!0$, Eq.~\ref{SHBspec}
yields a peak value of $E^2\, dN/dE$ at $E\!=\!E_p(t)$.
Observers usually report $E_p(t)$
at the peak's maximum, or its 
averaged value in a chosen time interval. For the 
approximate pulse shape
given by  Eq.~\ref{ICSPulse}, $E_p(t_{max})\!\approx\! 0.29\,E_p(0)$
while  $E_p\! \approx\! 0.23$  over the FWHM of a pulse.

For LGRBs, Eqs.~(\ref{PeakE}) give a good description of the observations, 
for $\epsilon_g\!\approx\! 2$ eV,  $\gamma_0\!\sim\! 10^3$, 
and $\delta_0\!\sim\!10^3 $ corresponding to the expected
$\theta\!\sim\! 1/\gamma_0.$
As their name reflects, SHBs typically have a larger $E_p$  than LGRBs. 
One reason is model-independent: SHBs are observed at smaller typical 
redshifts. In the CB model, according to Eqs.~(\ref{PeakE}), larger 
values 
of $(1+z)\, E_p$ may be due to a larger value of the 
typical $\epsilon_g$ and/or $\gamma_0$.
The glory of LGRBs being the very early light of a 
core-collapse SN, there are observational reasons to adopt 
$\epsilon_g\!\approx\! 2$ eV. For 
SHBs the ambient light may be the light of a companion star, or of an 
accretion 
disk. In the first case, there is also reason to adopt 
$\epsilon_g\!\approx\! ~2$ eV, which 
we will use for the sake of definiteness. 
The typical Lorentz and Doppler 
factors leading to the average value of $(1+z)\,E_p\!\sim\! 800$ keV
for the
well-measured cases in the table of SHB properties of Kann et al.~(2008) 
are:
\begin{equation}
\gamma_0[{\rm SHB}]\sim \delta_0[SHB]\sim 1400,
\label{gamshb}
\end{equation} 
somewhat larger than the typical values, 
$\gamma_0\!\sim\delta_0\!\sim\! 10^3$ for LGRBs.

\subsection{The width and time-lag of SHB pulses}

The peak time of $dN/dt$ for the pulse-shape given by 
Eq.~(\ref{ICSPulse}) is around $t=\Delta t\!=\! t_{tr}$, where
\begin{equation}
t_{tr}\approx {\sqrt{3}\, (1+z)\over \delta_0\, \beta_s}\, {R_{tr}\over c}
\approx {23\, {\rm ms}\over \beta_s}\, {10^{3}\over \delta_0}\, (1+z)\,
\sqrt{N_B\over 10^{48}}\,.
\label{ttr}
\end{equation} 
The full width of $dN/dt$
at half maximum is FWHM$\sim 2.38\, t_{tr}$ and its rise-time  
from half maximum to peak value is $t_{rise} \sim 0.59\, t_{tr}$
(FWHM$\sim 1.8\, t_{tr}$ and $t_{rise} \sim 0.7\, t_{tr}$
for the pulse shape used in DDD2004).
The pulse-shape 
is energy-dependent because of the exponential factor
in the pulse light curve,
\begin{equation}
E {d^2N\over dE}  \propto e^{-E/E_p(t)}\rightarrow 
          e^{-2\,E\, t^2/E_p(0)\, \Delta t^2}\,, 
\label{expf} 
\end{equation} 
and the rise time, peak time and FWHM of the ICS pulse are 
energy-dependent and decrease with increasing energy. In the CB model, 
this explains the observed time-lag effect in LGRB pulses which decreases 
with increasing energy. As can be seen from Eq.~(\ref{expf}), the time-lag 
is 
proportional to the width parameter, $\Delta t$, and decreases with 
increasing $E_p$, yielding a vanishingly small time-lag in SHBs
with a very small width and a large $E_p$. For 
$t^2>>\Delta t^2$, the pulse has a shape, 
\begin{equation} 
{d^2N\over dt\, dE} \propto (E\,t^2)^{-1}\, e^{-2\,E\, t^2
                    /E_p(0)\, \Delta t^2}=F(E\,t^2)\, . 
\label{Ets} 
\end{equation} 
This `$E\,t^2$' law is a test of ICS on a glory's light that is 
becoming more radially directed at increasing radii and times: $\langle 
1\!+\!cos\theta_{in}\rangle\rightarrow r_g^2/2\,r^2\sim  \Delta 
t^2/2\,t^2$ in 
Eq.~(\ref{ICSPulse}). This law, in its form as a correlation between the 
FWHM of 
collections of pulses and the energy interval at which they are measured 
($FWHM\sim E^{-1/2}$) has been known for a long time (e.g.,
Ramirez-Ruiz \& Fenimore et 
al.~2000). For 
well-measured single-pulse XRFs, Eq.~(\ref{Ets}) can be tested with 
precision, from X-ray to optical frequencies, both as an $(E,t)$ 
correlation 
and as a spectral form (DDD2007b). One example is XRF 060218, specifically 
discussed in De R\'ujula (2008). For SHBs we do not have enough information 
to test Eq.~(\ref{Ets}) and its underlying assumption that the source of 
the ambient light is localized close to, or around, the engine.

The measured redshifts of SHBs are relatively small, averaging $\langle 
z\rangle\sim 0.5$ compared with those of LGRBs; $\langle 
z\rangle\sim 2$. The widths of their pulses have been studied in more 
detail than their rise times, and range from 5 to 300 ms, with a broad 
peak at 50 ms (Nakar~2007, Kann et al.~2008). These numbers correspond to 
full widths at quarter maximum,  $FWQM\!\sim\! 3.5\,t_{tr}$ for the pulse 
shapes. The typical values $\gamma_0\!=\!\delta_0\!=\!1400$, $z\!=\!0.5$,
and $\Delta t\!=\!10$ ms in the observer's frame correspond to
$r_g\!=\! c\, \gamma(t)\, \delta(t)\, \Delta t/(1+z)\! \approx\! 4\times 
10^{14}$ cm in the progenitor's rest frame.
Inverting Eq.~(\ref{ttr}) and using  
$\gamma_0\!=\!\delta_0=1400$, we obtain $N_B[{\rm SHBs}]\!\sim\! 
10^{48}$ for $FWQM=50$ ms and 
$\beta_S\!=\!1/3$, 
respectively.  This is smaller than the typical $N_b[{\rm LGRBs}]\sim 
10^{50}$ by two orders of magnitude, probably because 
core-collapse SNe have much more available mass for potential 
CB-generating  accretion than any of the putative SHB progenitors.

\subsection {The peak luminosity and isotropic energy of SHBs}

Let $L$ be the luminosity of the source (an accretion disk or a massive 
companion). For a transparent or 
semi-transparent distribution of circumburst material,
it is approximately the glory's luminosity. The peak luminosity of a 
CB is at $t=t_{tr}$, the
time it becomes of transparent, and it is given by (DD2004),
\begin{equation}
(1 + z)^2\, L_p ={\delta_0^4\, \beta_s^2\, L\over 9} \, .
\label{Lpeak}
\end{equation}
In principle, Eq.~(\ref{Lpeak}) could be 
used to estimate the typical
$L$ of the ambient light of SHBs. 
However, the results for $L_p$ are often 
reported for binning
times much larger than the peak rise-times (e.g., Gehrels et al.~2006) and 
cannot be used to
estimate $L$. 
For an SHB with $N_{CB}$ prominent pulses of similar properties, the 
isotropic-equivalent energy, $E_{iso}$, is (DD2004):
\begin{equation}
E_{iso}={\delta_0^3\, L\, \beta_s\, N_{CB}\over 6\, c}\,
        \sqrt{\sigma_{_T}\, N_B \over 4\, \pi}=
        (1.2\times 10^{50}\, {\rm erg})\, N_{CB}\, \beta_s\, 
        \left({\delta_0\over 10^3}\right)^3\, 
      {L \over 10^{40}\, {\rm erg\, s^{-1}}}\, \sqrt{{N_b \over 10^{48}}}.  
\label{Eiso}
\end{equation}
The typical $E_{iso}$ of SHBs is a few $10^{50}$ ergs,
with a wide spread between $10^{48}\!-\!10^{52}$ erg 
(see e.g. Kann et al.~2008).
The typical values $\delta_0\!=\!1400$, $\beta_s\!=\!1/3$, $N_{CB}\!=\!3$
and $N_B\!=\!10^{48}$ yield the estimate, $L\!\sim\! 10^{40}\, {\rm erg\,  
s^{-1}}$, which is quite normal for the putative progenitors of SHBs.

\subsection{Correlations between prompt emission observables}

The pronounced dependence of LGRBs observables on the Doppler factor, 
that ranges over a broad domain, led to simple correlations between them 
(DD2000, DDD2004. DD2007a). Similar correlations between prompt emission 
observables are expected in  SHBs if they are dominated by a single 
class of progenitors. But the scarcity of data and lack of 
statistics on SHBs with secured redshift do not yet allow conclusive 
tests 
of such correlations.

\begin{figure}[]
\centering
\vspace{-1cm}
\epsfig{file=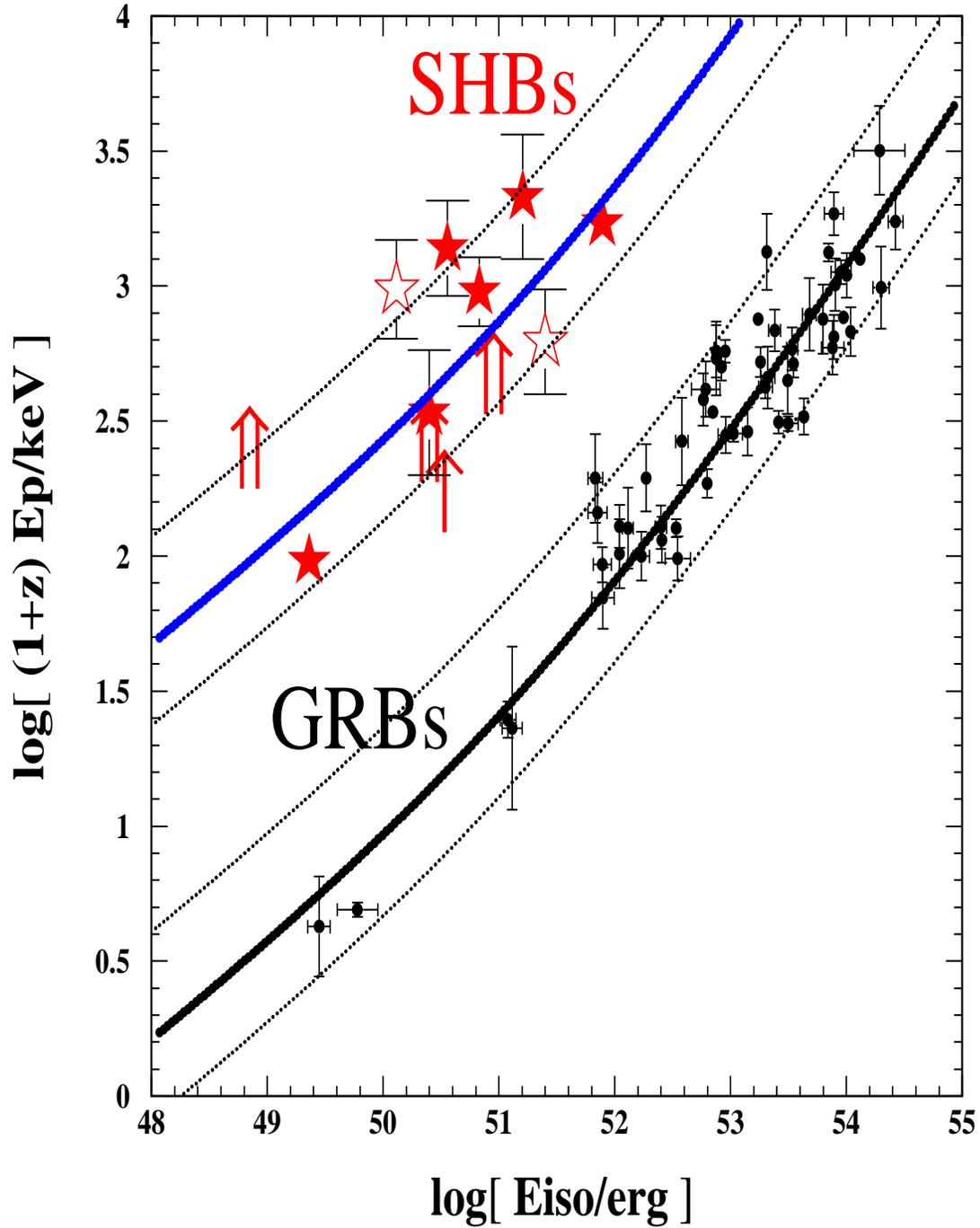,width=16.0cm,height=20.0cm}
\caption{
Comparison between the observed correlation $[E_p, E_{iso}]$
in LGRBs and SHBs and the CB model expectations for LGRBs (DDD2007a) and 
SHBs as given by Eq.~(\ref{EpEiso}).
}
\label{F5}
\end{figure}

The simplest CB model correlations for single pulses are
$\Delta t\! \propto\! E_p^{-1}$, and
$ E_{iso}\!\propto\! (1 + z)^2\, L_p^{4/3}$.
They are well satisfied for LGRBs (DD2004).
For other pairs of observables, the predicted correlations are slightly 
more elaborate (DDD2007a). For instance,
given that $(1 + z)\, E_p\! \propto\! \delta_0\, \gamma_0$
and  $E_{iso}\!\propto\!\delta_0^3$
one expects, $(1 + z)\,E_p\!\propto\! E_{iso}^{1/3}$. But that
is precise only for the low $E_p$ and $E_{iso}$ (large $\theta$)
cases, at the XRF limit of the LGRB
distribution. For $\theta\lsim 1/\gamma_0$,  
$(1+z)\, E_p\!\propto\! \gamma_0^2$ and  $E_{iso}\propto \gamma_0^3$   
implying that
$(1 + z)\, E_p\!\propto\! E_{iso}^{2/3}$. Since the observer's angle 
varies continuously, XRFs and
GRBs lie, in the $[(1+z)\, E_p,E_{iso}]$ log-log plane, close to a 
line 
whose slope varies smoothly
from 1/3 to ~2/3 (DDD 2007a), as shown in Fig.~\ref{F5}.
In Fig.~\ref{F5} we 
also show the limited observational information on the $[(1+z)\, E_p,\, 
E_{iso}]$ correlation for SHBs:
The results, like those for LGRBs, are 
fitted to a power-law varying from a 1/3 to a 2/3 slope (DDD2007a):
\begin{equation}
(1+z)\, E_p\approx  E_p^0\,
 ([E^\gamma_{iso}/E_0]^{1/3}+[E^\gamma_{iso}/E_0]^{2/3}), 
\label{EpEiso}
\end{equation}
with two parameters $E_p^0, \, E_0$.
They are almost
equally well fit by just the higher power. This is not surprising, for 
selection effects may imply that, so far, only the most energetic SHBs 
have been observed. 
SHBs viewed far off axis, are wider and should look 
like 
GRBs of low luminosity without an associates SN. 

The correlations between $E_p$ and other prompt observables (peak 
luminosity, peak rise-time, lag-time, and variability), which are 
satisfied for LGRBs (DDD2007a), are also very straightforward tests of the 
hypothesis of ICS for SHB produced by a single class of progenitors, but 
the data on SHBs are not precise enough to reach a conclusion at the 
moment.

\section{The synchrotron radiation afterglow}

As a CB plough through the ISM, its radiation 
ionizes the gas in front of it. In its rest frame, the 
ionized ISM particles impinge on the CB  with a Lorentz factor 
$\gamma$ and generate in it a turbulent magnetic field 
in equipartition with their energy density, 
$B\approx \sqrt{4\, \pi\, n\, m_p\, c^2}\, \gamma$.
The swept-in electrons are Fermi accelerated by the CB's turbulent
magnetic field and emit 
synchrotron radiation. The SR, isotropic in the CB's
rest frame, has a characteristic frequency, $\nu_b(t)$,
the typical frequency radiated by the
electrons that enter a CB at time $t$ with a relative Lorentz
factor $\gamma(t)$. In the observer's frame:
\begin{equation}
\nu_b(t)\simeq  {\nu_0 \over 1+z}\,
{[\gamma(t)]^3\, \delta(t)\over 10^{12}}\,
\left[{n\over 10^{-3}\;\rm cm^3}\right]^{1/2}
{\rm Hz}.
\label{nub}
\end{equation}
where $\nu_0\!\sim\! 8.5\times 10^{15}\, \rm Hz \simeq 35\, eV$.
The spectral energy density of the SR
from a single CB at a luminosity distance $D_L$  is given by (DDD2003a):
\begin{equation}
F_\nu \simeq {\eta\,  \pi\, R^2\,n\, m_e\, c^3\,
\gamma(t)^2\, \delta(t)^4\, A(\nu,t)\,
\over 4\,\pi\, D_L^2\,\nu_b(t)}\;{p-2\over p-1}\;
\left[{\nu\over\nu_b(t)}\right]^{-1/2}\,
\left[1 + {\nu\over\nu_b(t)}\right]^{-(p-1)/2}\,,
\label{Fnu}
\end{equation}
where $p\sim 2.2$ is the typical spectral index\footnote{The normalization
in Eq.~(\ref{nub}) is only correct for $p\!>\!2$, for otherwise the norm
diverges. The cutoffs for the $\nu$ distribution are time-dependent,
dictated by the acceleration and SR times of electrons and their `Larmor'
limit. The discussion of these processes being complex (DD2003a, DD2006),
we shall satisfy ourselves here with the statement that for $p\!\leq \!2$
the AG's normalization is not predicted.} of the Fermi accelerated
electrons, $\eta\!\approx\!1$ is the fraction of the impinging ISM 
electron energy that is synchrotron re-radiated by the CB, and $A(\nu, t)$ 
is the  attenuation of photons of observed frequency $\nu$ along the
line of sight through the CB, the host galaxy (HG), the intergalactic 
medium (IGM) and the Milky Way (MW):
\begin{equation}
A(\nu, t) = {\rm
e^{-(\tau_\nu(CB)+\tau_\nu(HG)+\tau_\nu(IGM)+\tau_\nu(MW))}.}
\label{attenuation}
\end{equation}
The opacity
$\tau_\nu\rm (CB)$ at very early times, during the fast-expansion phase of
the CB, may strongly depend on time and frequency. The opacity of the
circumburst medium [$\tau_\nu\rm (HG)$ at early times] is affected by the
GRB and could also be $t$- and $\nu$-dependent.  
The opacities
$\tau_\nu\rm (HG)$ and $\tau_\nu\rm (IGM)$ should be functions of $t$ and
$\nu$, for the line of sight to the CBs varies during the AG observations,
due to the hyperluminal motion of CBs. These 
and the dependence of
the synchrotron AG on $\nu_b(t)$ may generate complex 
chromatic behaviour in the AGs (DDD2007b).

The Swift X-ray bands are above the characteristic frequency $\nu_b$ in
Eq.~(\ref{nub}) at all times. It then follows from Eq.~(\ref{Fnu}) that
the {\it unabsorbed} X-ray spectral energy density has the form:
\begin{equation}
F_\nu \propto R^2\, n^{(p+2)/4}\,
\gamma^{(3p-2)/2}\, \delta^{(p+6)/2}\,  \nu^{-p/2}=
R^2\, n^{\Gamma/2}\,
\gamma^{3\,\Gamma-4}\, \delta^{\Gamma+2}\, \nu^{-\Gamma+1}\, ,
\label{Fnux}
\end{equation}
where we have used the customary notation
$dN_{\gamma}/dE\!\approx\!E^{-\Gamma}$.
The ISM density in globular clusters and in the surrounding
medium, if they are far off galactic center, may be low enough such that 
the 
optical band is also above the bend frequency $\nu_b$, and then the 
optical AG 
is also described by Eq.~(\ref{Fnux}) and the AG is achromatic  all the 
way between the optical to the X-ray band with a spectral 
index $\beta_{OX}\simeq 1.1$.    

\subsection{The early time SR afterglow}
During its early-time emission, when  both $\gamma$
and $\delta$ stay put at their initial values $\gamma_0$ and $\delta_0$,
Eq.~(\ref{Fnu}) yields an early time light curve
$F_\nu \propto e^{-\tau_{_W}}\, R^2\, n^{(1+\beta)/2}\,\nu^{-\beta}.$
If the progenitor is embedded inside a globular cluster or a superstar 
cluster which blows a constant wind into the ISM or IGM  
with a density  profile $n\!\propto\! 1/r^2\!\sim\! 1/t^2$,   
the early time SR light curve as given by Eq.~(\ref{Fnu}) has the form, 
\begin{equation}
F_\nu \propto  {e^{-a/t}\,
t^{1-\beta}\, \nu^{-\beta}
\over t^2+t_{exp}^2}\,, 
\label{SRP}
\end{equation}
until the CB reaches the constant ISM or IGM density.

\subsection{Jet breaks, missing breaks and asymptotic decay}
As it ploughs through the ionized ISM, the CB
gathers and scatters its constituent ions, mainly protons. These 
encounters
 are `collisionless' since, at about the time it becomes transparent to
radiation, a CB also becomes `transparent' to hadronic interactions
(DD2004). The scattered and re-emitted
protons exert an inward pressure on the CB, countering its expansion.
In the approximation of isotropic re-emission in the CB's
rest frame and a constant ISM density $n\!\sim\!n_e\!\sim\!n_p$,
one finds that within a minute or so of observer's time $t$,  
typical SHB generating CBs
of baryon number $N_b\!\sim\! 10^{48}$ reach an approximately constant
`coasting'  radius $R\!\sim\!10^{14}$ cm, before they finally
stop and blow up, after a journey of years of observer's
time. During the coasting phase, and in a constant density ISM,
$\gamma(t)$  obeys (DDD2002a, DDD2006):
\begin{equation}
({\gamma_0/ \gamma})^4 +
(2\,\theta^2\,\gamma_0^2\,(\gamma_0/\gamma)^2
= 1+2\,\theta^2\,\gamma_0^2+t/t_0\,,
\label{decel}
\end{equation}
with
\begin{equation}
t_0={(1+z)\, N_{_{\rm B}}\over 8\,c\, n\,\pi\, R^2\, \gamma_0^3}.
\label{break}
\end{equation}
As can be seen from Eqs.~(\ref{decel},\ref{break}), $\gamma (t,t_0,\theta, 
\gamma_0)$,  
and consequently also $\delta$,
change little as long as $t\!<\! t_b\!\equiv\![1+2\,\gamma_0^2\, 
\theta^2]\, t_0 $, where,
\begin{equation}
t_b= (1300\,{\rm s})\, [1+2\,\gamma_0^2\, \theta^2]\,(1+z)
\left[{\gamma_0\over 10^3}\right]^{-3}\,
\left[{n\over 10^{-3}\, {\rm cm}^{-3}}\right]^{-1}
\left[{R\over 10^{14}\,{\rm cm}}\right]^{-2}
\left[{N_{_{\rm B}}\over 10^{48}}\right] \!,
\label{tbreak}
\end{equation}
and they approach a power-law decay, 
$\delta\propto \gamma\!\sim\! t^{-1/4}$ for  $t\!>\!t_b$. 
The slow change in $\gamma$ and $\delta$ for $t\lsim t_b$
produces the  "plateau" phase of  canonical afterglows.
Their asymptotic power-law decay  for $t\gg t_b$, when  
inserted in  Eq.~(\ref{Fnux}), yields the asymptotic power-law decay 
(DDD2007b),
 \begin{equation}
F_\nu(t)\sim  t^{-\beta-1/2}\, \nu^{-\beta}=t^{-\Gamma +1/2}\, 
             \nu^{-\Gamma+1},   
\label{asympt1}
\end{equation}
where $\beta\!\equiv\!\Gamma\!-\!1\!=\!p/2\!\sim\! 1.1$. 
Eq.~(\ref{asympt1}) is valid as long as the
ISM has an approximately constant density.  
The gradual transition (`break') of the AG from the 
plateau phase 
to a power-law decay at $t\gsim t_b$ takes place when the CB has swept 
in a mass comparable to its initial mass. This CB model bend/gradual 
break is different from the 
achromatic break predicted by the conical Fireball model   
when the ejecta decelerate to a point when the observer begins to see the 
entire opening angle 
of a conical ejecta (Rhoads~1997,1999).

Density variations complicate the simple shape 
of AGs and will not be discussed here, except for noting that 
for an ISM with an approximate $\!\sim\! 1/r^2$ density profile, such 
as in a globular cluster, or in a galactic halo with an isothermal sphere 
density profile, which CBs may reach late in their motion, or 
in a windy environment created 
around superstar clusters by their blowing winds, 
the predicted X-ray and optical AG has a simple power-law form:
\begin{equation}
F_\nu(t)\sim  t^{-\beta-1}\, \nu^{-\beta}=t^{-\Gamma}\,
             \nu^{-\Gamma+1}\, .
\label{asympt2}
\end{equation}
In ordinary LGRBs, $t_b$ is usually larger than a few hundreds of seconds   
but, in intrinsically bright LGRBs, in particular in those which  happen 
to take place in a dense ISM environment, the break in the X-ray
AG may occur before the end of the prompt emission.  Such breaks may be 
hidden under the much brighter ICS emission  and only the power-law
decaying tail of the AG is observed (DDD2008b). In  SHBs, the low density 
environment of globular clusters, in particular those 
with a large offset from the galactic center, yield 
relatively large $t_b$ values as can be seen from  Table II. 
For SHBs in young superstar clusters with a strong wind,
the AG has the simple power-law decay given by 
Eq.~(\ref{asympt2}) with no jet break.

\section{The extended soft emission component}

Globular clusters (GCs) are concentrations of 
$\sim\! 10^4$ to $\! \!10^8$ 
stars in and around galaxies.
(e.g. the GC 037-B327 in M31 
has a mass $M_{GC}\!=\!(3\!\pm\! 0.5)\times 
10^7\, M_\odot$,  Ma et al.~2006), 
They are the environment wherein a considerable 
fraction of low- mass X-ray binaries (compact stars acreeting from a 
companion), milli-second pulsars (neutron stars having been spun-up by 
accretion) and neutron star binaries are found. Since these are some of 
the putative progenitors of SHBs, it is natural to test whether some 
features of these bursts, other than their distribution relative to the 
host-galaxy center (Troja et al.~2007, Kann et al.~2008), are also 
compatible with a GC location. 
GCs have typical core radii of $r_c\simeq 1$ pc.
Sometime they have an intermediate mass black hole at their center. 
Relatively large GCs have a core-luminosity density of $O(10^5\, L_\odot\, 
{\rm pc^{-3}})$ and a total luminosity $L_{GC}$ of order $10^6\, 
L_\odot\!\sim\!  4\times 10^{39}\, {\rm erg\, s^{-1}}$.

The more recently discovered (Arp \& Sandage~1985) superstar clusters 
(SSCs) are star concentrations similar to GCs in their sizes, but denser, 
more massive, and more active. A good fraction of SSCs may be 
gravitationally bound and constitute a proto-globular cluster, thus the 
alternative denomination young globular clusters. Many active hosts 
harbour SSCs, including interacting galaxies, starburst galaxies and 
star-formation regions in normal spirals. 
Dust often obscures SSCs, which are not easy to see at optical 
frequencies, 
and have only recently been studied in detail with the HST, mostly in the 
Antennae galaxies (Whitmore et al.~1995), in M82 and in NGC 2553. The 
sizes of observed SSCs range from 1 to 6 pc, their masses from a few
$10^4$ to 
$10^6\, M_\odot$ and their luminosities from $10^{40}$ to 
$10^{42.5}\, {\rm erg\, s^{-1}}$, nearly $10^9\, L_\odot$ 
(Melo et al.~2005 and references therein). Even at small redshifts, 
active 
SSCs are not a rare phenomenon; they must have been rather common in the 
past. Because of their enhanced stellar-evolution activity, SSCs are even 
more natural hosts than GCs for the putative progenitors of SHBs.

We shall argue that the extended soft component observed in a good 
fraction of SHBs is due to ICS of the light of a superstar cluster by the 
CBs crossing it, after they have left the close-by luminous environment 
generating the SHBs prompt peaks, or SR in young superstar clusters with 
a very large ($n\!\sim\! 10^3\!-\!10^6\, {\rm cm^{-3}}$) ISM density. This 
is supported by the typical duration and isotropic energy of ESECs 
which are 
compatible with the said hypothesis. Concerning the ESECs spectrum, the 
data are only sufficient to conclude that the hypothesis is consistent.

\subsection{The duration of an ESEC}
For the measured average SHB redshift, $z\!\sim\!0.5$, 
and with 
the typical values, $\delta_0\,\gamma_0\!\sim\! 1400$, the distance 
travelled 
by a CB in the typical $t_{ESEC}\!\sim\! 100$ s duration of an ESEC is 
$x\!\sim\!\gamma_0\,\delta_0\, c\, t_{ESEC}/ 
(1\!+\!z)\!\approx\! 1.2$ pc, coincident with the typical core radius 
of a GC or SSC. This straightforward understanding of ESEC durations is 
independent of whether the mechanism generating the radiation is ICS or 
SR, provided the sources are CBs, moving relativistically, and not 
significantly decelerating in this short interval of observers time.

\subsection{The ICS contribution to the ESEC}

As a CB moves through the light of the a host star cluster  
its ICS produces  a light curve which 
traces the density of the light along the CB trajectory.  
This light is a sum of a smooth background light
from all the stars in the cluster plus light peaks from the passage near 
very 
luminous stars. For simplicity, consider  
the contribution to the ESEC light curve from ICS of background light 
by a CB ejected near the center of an SSC. 
It is obtained simply by replacing   
the typical energy $\epsilon_g$ 
and density $n_g(t)$ of the glory photons 
in the expression describing the prompt SHB with, respectively,  the
typical energy $\epsilon_{c}$ and the density 
$n_c(t)$ of the GC background photons, where
$n_c(0)\!=\!3\, L_{ssc}/4\, \pi\, c\, \epsilon_c\ r_c^2$
and $L_{ssc}$ is the SCC luminosity. 
The resulting ESEC light curve is given by 
Eqs.~(\ref{ICSPulse}) and (\ref{SHBspect})
with $\Delta t$ replaced 
by $\Delta t_c\!=\! r_c\, (1+z)/ c\,\gamma_0\, \delta_0$.
The isotropic-equivalent energy of the ESEC is then given by,
\begin{equation}
E_{iso}[ESEC]\approx \sigma_{_T}\, N_b\, N_{CB}\,\gamma_0\, \delta_0^3\,    
               {3\, L_{ssc}\over 4\, c\, r_c}
            \sim (10^{49}\, {\rm erg})\, {N_b\over 10^{48}}\, 
             {N_{CB}\over 4} \, {pc\over r_c} {\gamma_0\, \delta_0^3\over 
              1400^4}\, {L_{ssc}\over 10^8\, L_\odot}\, .
\label{EISOESEC}
\end{equation}
The result of Eq.~(\ref{EISOESEC}) is of the observed order of magnitude, 
but has more sources of
accumulated uncertainty and variability than other CB-model results.

The ESEC after transparency time, for a spherical SSC, 
is a decreasing 
function of time from the beginning, if the progenitor is in the front 
hemisphere, and the CBs' distance from the GC increases. Otherwise, the 
ESEC light curve first increases and then decreases with time. The data 
are 
consistent with these expectations, but so far insufficient to test them 
in detail.

\subsection{The SR contribution to the ESEC} 
Some young superstar clusters have a very large ISM density, 
$10^3\lsim ~n~ \lsim 10^6\, {\rm cm^{-3}}$. Consequently, 
the emission rate of SR radiation from 
CBs moving inside such SSCs,  
which is proportional to the ISM density (see Eq.~(\ref{Fnu})),
is very intense and  relatively hard with a spectral index  
$\beta\simeq 1/2$, because 
$\nu_b\!\propto\! n^{1/2}$ (see Eq.~(\ref{nub})) is well above 
the X-ray band. At early time it is given by Eq.~(\ref{SRP})
with $\beta\simeq 1/2$.
As soon as the CB escapes out of a  dense
core of a superstar cluster, the density 
decreases fast like 
$1/r^2$, 
and the light curve decreases rapidly and becomes  softer
with its spectral index  increasing to $\beta\!\approx \!1.1\,,$
\begin{equation}
F_\nu \propto  {e^{-a/t}\,
t^{1/2}\, \nu^{-1/2}
\over t^2+t_{exp}^2}\rightarrow  t^{-2.1}\, \nu^{-1.1}.
\label{SRSSC}
\end{equation} 
Eq.(\ref{SRSSC}) is valid when the CBs encounter  a very large density 
inside 
the SSC such that the CBs reach their coasting radius,
$R_{cb}\!\propto\! N_b^{1/3}\, n^{-1/3}\, \gamma_0^{-2/3}$, well inside 
the SSC and $t_{exp}\!\ll\! \Delta_{ssc}$.

\section{Comparison between SHB afterglows and their CB model
description}

\subsection{General}

To date, Swift has detected over 30 SHBs, localized them through their 
$\gamma$, X-ray and UVO emissions and followed them until they faded into 
the background, usually within less than a day or two. A few deep 
observations of the X-ray emission of several SHBs at later times were 
made with Chandra. Several additional SHBs were localized by the Inter 
Planetary Network (IPN) to much larger error boxes and a couple more by 
HETE. Beside the Swift UVO observations, there have been many optical 
follow-up measurements of SHBs by ground-based optical telescopes 
including some of the largest ones. In order to demonstrate the success of 
the CB model to explain SHBs, we have limited the detailed comparison 
between theory and observations to all SHBs (15) whose X-ray and/or 
optical AG were well sampled. The {\it a-priori} unknown parameters are 
the 
number of CBs, their ejection time, baryon number, Lorentz factor and 
viewing angle, and the distributions of the glory's light and the ISM 
density along the CBs' trajectory.

In more than 2/3 of the SHBs detected by Swift, no ESEC was detected and 
only in few SHBs were its light curve and spectrum  measured. Several
SHBs have no measured redshift. So far no absorption or emission lines 
were detected in the afterglow of SHBs due to their faintness (Stratta et 
al.~2007). Their redshifts listed in Table I were determined from their 
host galaxies or from the nearest bright galaxy or galaxy cluster if no 
galaxy or only extremely faint galaxies were found in their XRT error 
circles. But, the lack of secured redshifts and well measured ESECs does 
not 
prevent other critical tests of the CB model predictions, which are 
insensitive to the unknown redshift and the detailed behaviour 
of the ESEC.

We have assumed quite generally an isothermal sphere density profile, $ 
n\propto 1/(r^2\!+\!r_c^2)$, for both the star clusters and galactic 
halos, with $r_c\!\sim\!  O(1\,pc)$ for GCs and SCCs, and $r_c\!\sim\! 
O(10\,kpc)$ for galactic halos, until being taken over by a 
constant-density ISM or IGM, respectively. This density profile describes 
also the windy density profile outside young superstar clusters or SN 
remnants, created by the strong winds from SSCs or the 
progenitor star of core collapse SNe, respectively.

Because of the short duration of SHBs and the fast initial expansion of 
the CBs, we have assumed that by the time they have escaped from  
the 
globular/superstar cluster, they have merged into a single CB. This 
reduces dramatically the number of fitted parameters without affecting the 
quality of the fits. Flares during the fast decay phase of the extended 
soft emission component (ESEC) or the afterglow phase due to late CB 
ejections are superimposed on the smooth light curve. To demonstrate that 
the CB model correctly describes all of the observed features of the Swift 
X-ray observations, it suffices to include in the fits only the latest 
observed pulses or flares during the ESEC. This is because the last 
exponential factor in Eq.~(\ref{expf}) suppresses very fast the relative 
contribution of earlier pulses by the time the data sample the later 
pulses or flares.

The X-ray  light curves
reported in the Swift/XRT GRB lightcurve repository (Evans et al.~2007) 
were fitted with use of Eq.~(\ref{ICSPulse}) for ICS pulses and 
Eq.~(\ref{Fnu}) for their synchrotron contribution, with an early time 
behaviour as given by Eq.~(\ref{SRP}). 

Below we summarize the observations and CB model description of all the 
SHBs  with well sampled X-ray light curve, reported in the Swift/XRT 
GRB lightcurve repository (Evans et al.~2007). 
and/or optical light curve.  The CB 
model predictions are based on Eq.~(\ref{ICSPulse}) for the 
prompt emission and the fast decay phase, 
Eq.~(\ref{Gamm}) for the evolution of the spectral index during the 
decay of the prompt emission or Eq.~(\ref{HR}) for the hardness ratio,
and Eq.~(\ref{Fnu}) for the SR afterglow. 
The hardness ratio during the SR dominated phase was fitted by a 
constant. The best fit parameters used in their CB model description are 
reported in Table II. Also reported there are the CB model fit parameters 
of the flares (subscript f) superimposed on the smooth XRT and/or optical 
light curves. Due to their faintness, in most cases the AGs of SHBs are 
not well measured beyond the plateau phase. This does not allow a reliable 
estimate of $p$ or the density profile from the late temporal decay of 
the AG. In such cases we have fixed $p$ to have its canonical 
theoretical value, $p\!=\!2.2$, and indicated that by (2.2) in Table II.

\subsection{Case Studies}

\begin{figure}[]
\centering
\vspace{-1cm}
\vbox{
\hbox{
\epsfig{file=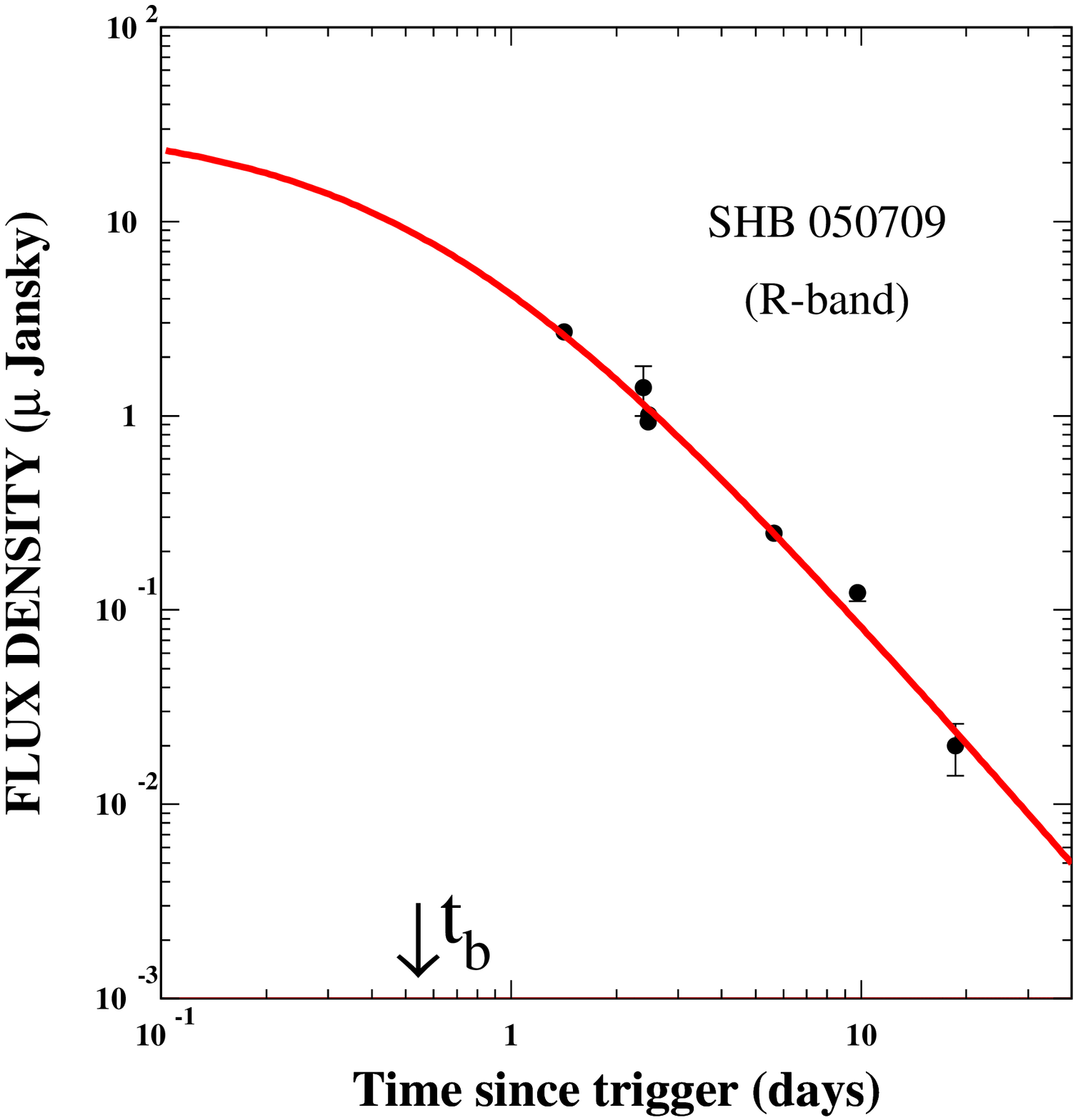,width=8.0cm,height=6.0cm}
\epsfig{file=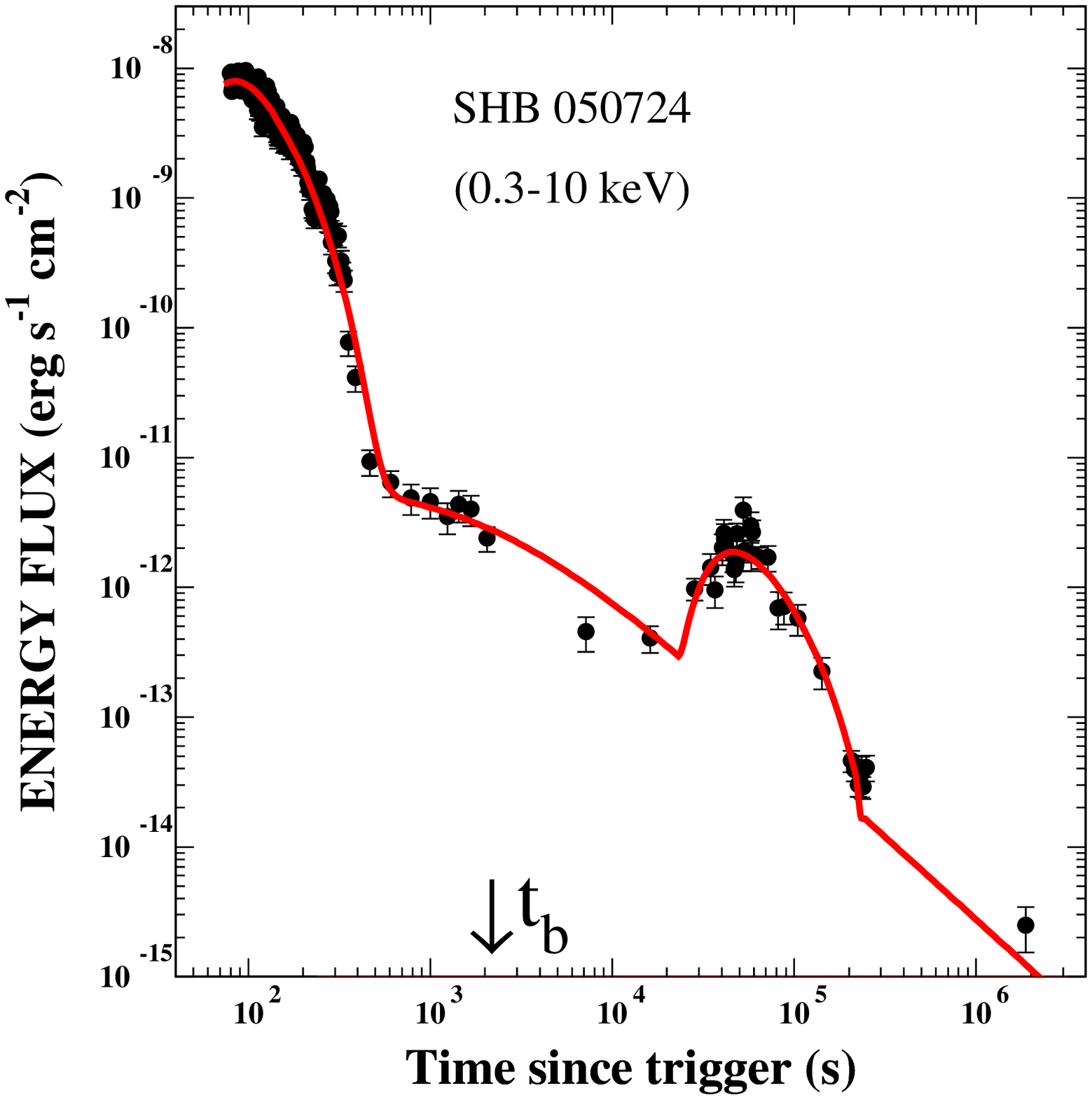,width=8.0cm,height=6.0cm}
}}
\vbox{
\hbox{
\epsfig{file=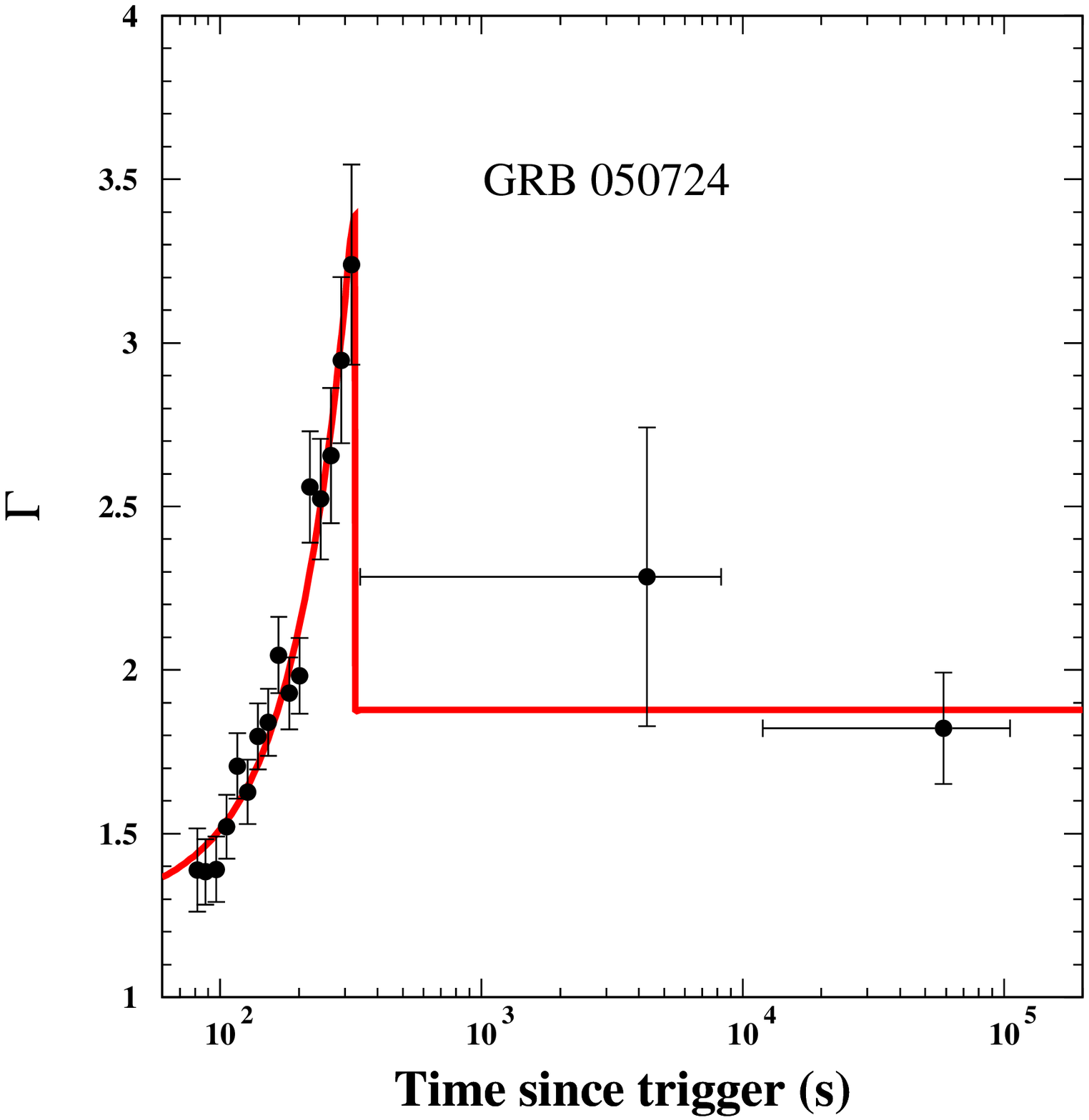,width=8.0cm,height=6.0cm}
\epsfig{file=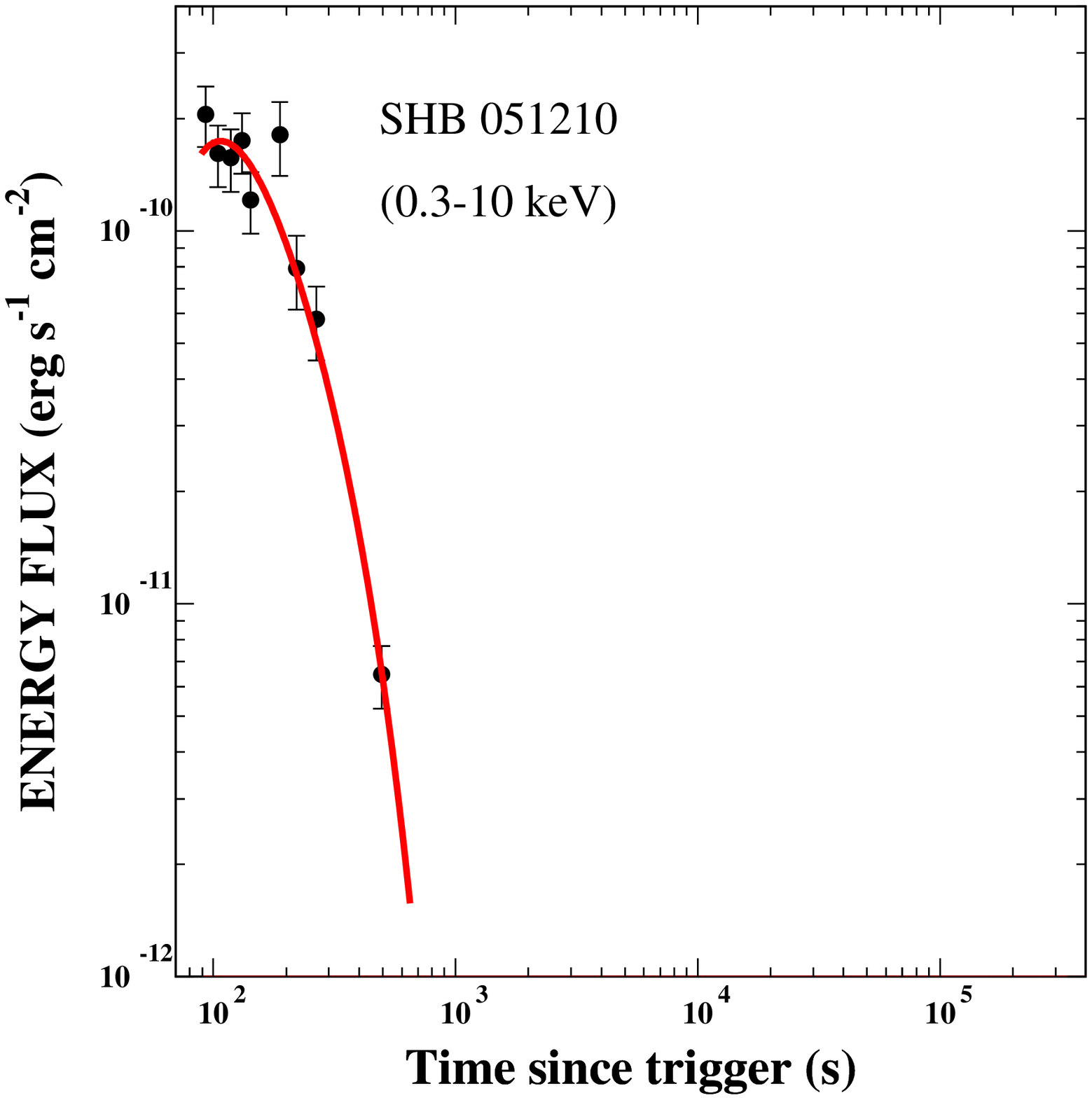,width=8.0cm,height=6.0cm}
}}
\vbox{
\hbox{
\epsfig{file=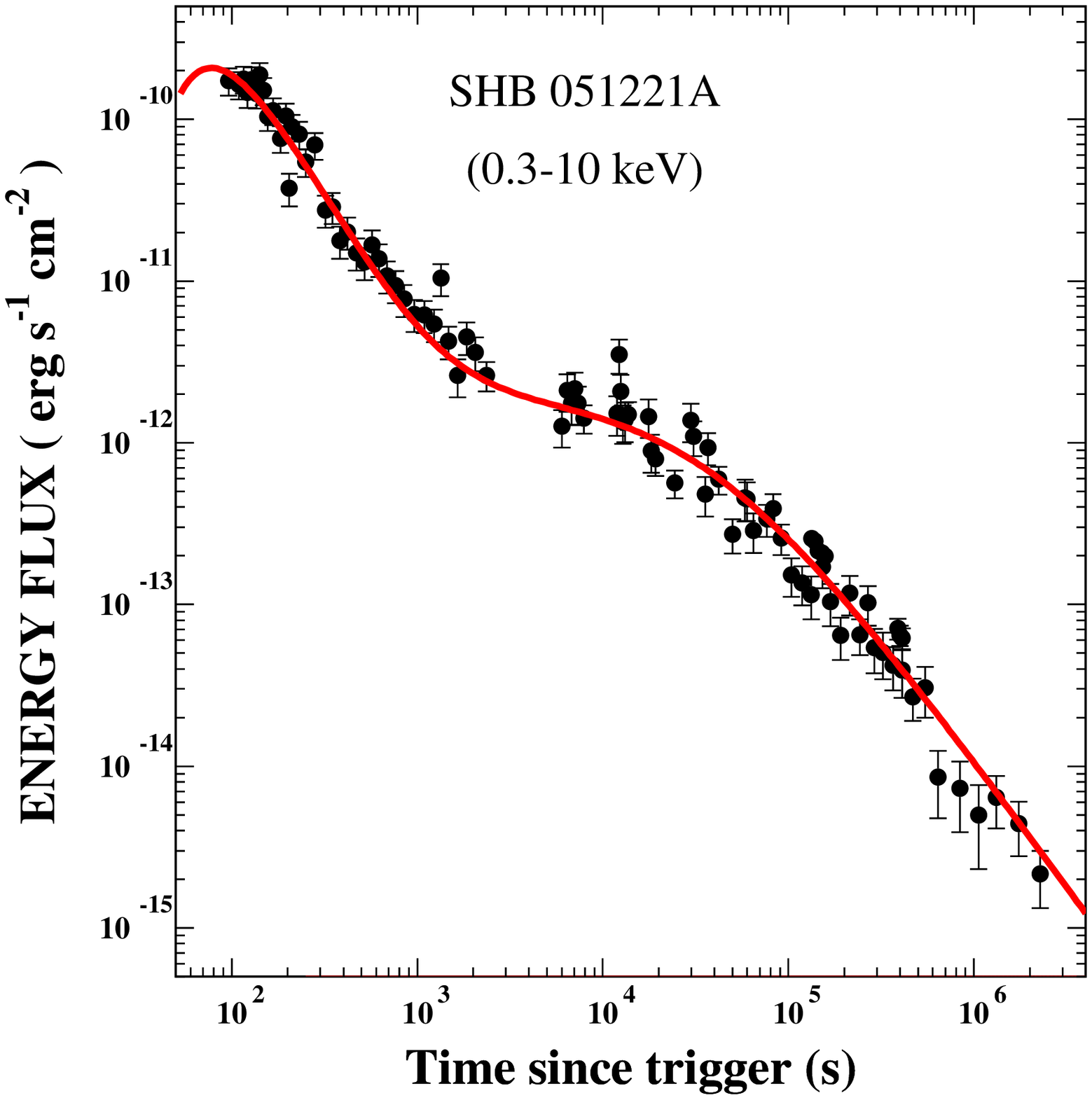,width=8.0cm,height=6.0cm}
\epsfig{file=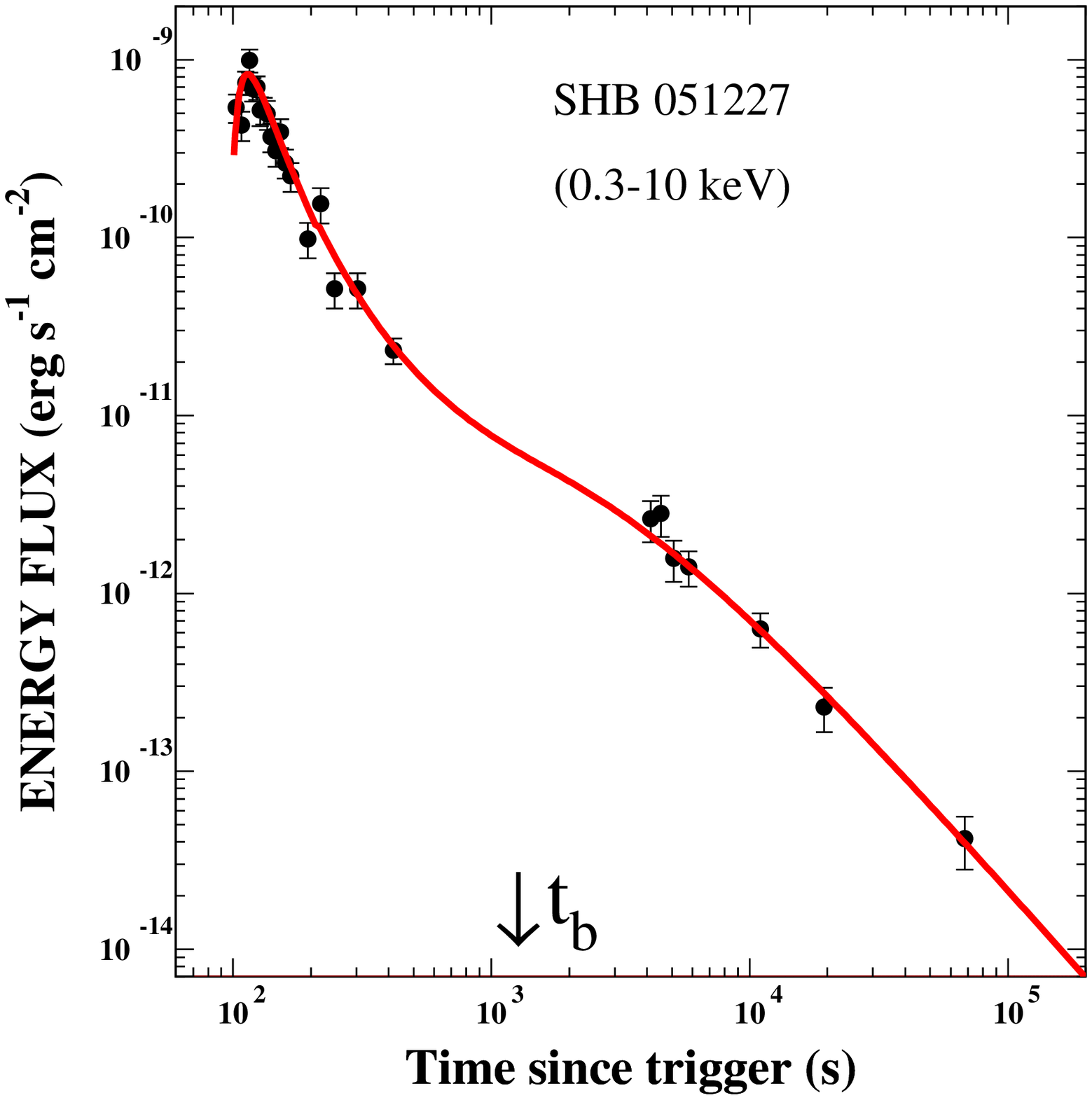,width=8.0cm,height=6.0cm}
}}
\caption{
Comparisons between SHB observations 
and CB model predictions for
{\bf Top left (a):} The R-band light curve of SHB 050724
(data from Watson et al.~2006).
{\bf Top right (b):} The XRT/Chandra light curve of SHB 050724
{\bf Middle left (c):} The evolution of the photon spectral index of 
GRB 050724 (data from Zhang et al. 2007).
{\bf Middle right (d):} The XRT light curve of SHB 051210.
{\bf Bottom left (e):} The XRT light curve of SHB 051221A.
{\bf Bottom right (f):} The XRT light curve of SHB 051227.
}
\label{F1}
\end{figure}

\noindent
{\bf SHB 050709.}\\
{\bf a. Observations:}
This SHB, which was localized by HETE (Vilasenor et al.~2005), was 
relatively very faint ($E_{iso}\!\sim\! 2.4\times 10^{48}$ erg). It had a 
multi-spike peak with $T_{90}\!=\!70\pm 10$ ms in the 30-400 keV band, 
$T_{90}\!=\!220\pm 50$ ms in the 2-25 keV band, and showed a fast hard to 
soft 
spectral evolution with $E_p\!\sim\! 86 \pm 16$ keV and a photon spectral 
index $\Gamma\!=\!0.82\pm 0.13$.  A soft extended emission component was 
detected 24 s after burst with $T_{90}\!=\!130\pm 7$ s, $\Gamma\!=\!1.98 
\pm 0.18$, and a fluence much larger than that of the short burst, unlike 
what is seen in the giant flares of SGRs. It was the first SHB for which an 
optical AG was discovered and was used to localize it in a star forming 
dwarf galaxy at redshift $z=0.16$ (Hjorth et al.~2005a, Fox et al.~2005, 
Covino et al~2006) at a projected distance of $\sim 3.8$ kpc from its 
center. Its X-ray AG was detected by Chandra (Fox et al.~2005). Its 
optical light curve as measured by Watson et al.~(2006)  is shown in 
Fig.~\ref{F1}a.

\noindent
{\bf b. Interpretation:} 
The short burst shows all the properties expected from ICS of glory 
light by a jet of CBs with a relatively small baryon number 
emitted by the source. The spectral index of the extended soft 
emission component (ESEC), $\Gamma=1.98 \pm 0.18$,  the lack of 
spectral evolution during the ESEC and the duration of the ESEC
are consistent with  those expected from SR ($\Gamma\sim 2.1$) 
of CB propagating in a $r_c\! \sim\! 1\, pc$  core of a globular cluster
and $t\! \sim\! (1+z)\, r_c / c\, \gamma_0, \delta_0 \sim 120$ s).
The optical afterglow is well explained  by a CB propagating out of 
the dwarf galaxy with an isothermal sphere  density distribution 
$n\!\propto\! 1/(r^2+r_0^2)$, as demonstrated in Fig.~\ref{F1}a.

\noindent
{\bf SHB 050724.}\\ 
{\bf a. Observations:} This burst at redshift $z=0.257$ was 
studied in detail in Campana et al.~2006b, Grupe et al.~2006 and 
Malesani et al.~2007. The BAT on board Swift triggered on the burst 
at 12:34:09 UT on 2005, July 24. The burst had $T_{90}\!=\!3.0\pm 1.0$ s, 
but most of the energy of the initial SHB was released in a hard spike 
with a 
duration of 0.25 s. The bulk of the burst energy was not emitted 
in the short initial spike but in an extended soft emission 
component with $\Gamma\!=\!2.5\!\pm\! 0.2$ which lasted $\!\sim 150$ s.
Swift's XRT began observing the afterglow 74 s 
after the BAT trigger. The Chandra X-ray observatory performed two 
observations, two days  and about three weeks after 
the burst. The complete X-ray light curve is shown in 
Fig.~\ref{F1}b. It has the canonical shape observed in long GRBs, 
namely, a rapid decay with a fast spectral softening ending with a 
sharp transition to a plateau phase with a much harder power-law 
spectrum, $\Gamma=1.79\pm 0.12 $ as shown in Fig.~\ref{F1}c. The AG 
steepens gradually into a late power-law decay. A large flare 
superimposed on the canonical light curve occurred around 50 ks after 
burst with a fluence of $\!\sim\! 7\%$ of that of the prompt burst. The 
flare has been detected also in the optical and NIR bands (e.g., 
Malesani et al.~2007). Spectral analysis of the XRT data (Campana 
et al.~2006b) showed no evolution during the afterglow phase, 
including the large late flare. Spectral analysis of the Chandra 
observations from the fading tail of this flare confirmed this 
result (Grupe et al.~2006). The burst took place 2.5 kpc (in 
projection) from the center of an elliptical host galaxy
(Malesani et al.~2007).

\noindent
{\bf b. Interpretation:} The CB model X-ray light curve 
of SHB 050724 and its spectral index `light curve'  
are shown in Figs.~\ref{F1}b,c. The early 
time emission is described by ICS of the progenitor's 
glory light, the ESEC
by ICS of the light of the core of the assumed superstar cluster   
or a globular cluster  environment.
The fast 
decay of the X-ray light curve and the rapid spectral softening took 
place when the CB was moving away from  the quasi-isotropic light 
distribution in 
the dense stellar environment of the progenitor  into the ISM of 
its elliptical host galaxy. As can be seen in Figs.~\ref{F1}b,c, this fast 
decay and spectral softening stopped simultaneously when the AG was 
taken over by the SR from the decelerating CB in an ISM of a
constant density. 
As expected, apart from normalization, the late-time SR afterglow is 
similar in shape to the SR afterglow of LGRBs. Also the late flare 
superimposed on the canonical light curve is similar to those observed in 
many LGRBs. It was produced by enhancement of the emitted SR when the CB 
encountered a density bump in its voyage through the host's ISM. The 
Chandra data show that the canonical AG continued to decay after the 
flare with the same slope and the same spectral index, $\beta_X\!=\!0.79 
\pm 0.15$. As 
shown, in Fig.~\ref{F1}b, the complete XRT light curve is well fit by the 
CB model. Moreover, the CB model relation 
$\alpha\!=\!\beta_X\!+\!1/2\!=\!p/2$ 
is well 
satisfied. The temporal behaviour of the canonical AG was best fit with 
$p\!=\!1.56$, implying an unabsorbed spectral index, 
$\beta_X\!=\!p/2\!=\!0.78$, and 
an asymptotic power-law decay with a power-law index, 
$\alpha\!=\!\beta_X\!+\!1/2\!=\!1.28$, in agreement with those observed.

The elliptical host galaxy of SHB 050724 was argued to 
provide strong support for a neutron star merger origin of this 
SHB. But, it was pointed out that neutron star mergers do not 
produce the late accretion episodes needed in the standard folklore to 
power a late central activity which could produce the large flare 
around 50 ks after burst (Grupe et al.~2006). In the CB model, a 
late flare with a typical SR spectrum and little spectral 
evolution is produced by density bumps along the CB trajectory in 
the ISM. Such flares neither rule out nor support any specific origin of 
the SHB.

\noindent
{\bf SHB 051210.}\\
{\bf a. Observations:} 
This was a faint two-spiked SHB localized by Swift.
It had $T_{90}\!=\!1.27\pm 0.05$ s and no soft extended
emission  was detected (La Parola et al.~2006). 
The BAT spectrum was fit with a power-law with a photon spectral index 
$\Gamma\!=\!1.1\!\pm\!0.3$.  The XRT started observation 79 s after the 
BAT 
trigger. It detected a rapidly decreasing  
light curve, shown in Fig~\ref{F1}d, with a small flare superposed around 
134 s. A possible
host galaxy was discovered
within the XRT error circle
(Bloom et al.~2005a), but no emission or absorption lines were 
detected. No optical afterglow was detected.

\noindent
{\bf b. Interpretation:}
The spectrum of the burst is consistent with ICS of glory light
($\Gamma\!\sim\! 1$ for $E \!\ll\! E_p$). 
The fast declining XRT light curve was fit 
as  the tail of an
ICS flare.  The data are insufficient for a critical test 
of the CB model interpretation.

\noindent
{\bf SHB 051221A.}\\ 
{\bf a. Observations:}
This burst was discussed in detail in
Soderberg et al.~2006a and Burrows et al.~2006.
It was an intense, multi-spiked burst localized by Swift, 
The burst was also detected by 
Konus-Wind, Suzaku, INTEGRAL and RHESSI and its late afterglow by Chandra. 
The burst had $T_{90}\!=\!1.4\!\pm\!0.2$ s, $\Gamma\!=\!0.92\!\pm\! 0.13$,
and showed no signs of extended  emission. 
The Swift/XRT observations of its X-ray light curve began 
88 s  after the Swift/BAT trigger. Its late time X-ray emission 
was also detected by Chandra.
The measured X-ray light curve (Fig.~\ref{F1}e) of SHB 051221A  shows the 
canonical behaviour of X-ray light curves of LGRBs (Nousek et al.~2005)
with an asymptotic decay index $\Gamma\!\approx\! 2.$
Its afterglow was detected also in the NIR, and optical bands.
and was localized near the center ($\!\sim\! 1$ kpc in projection)  in a 
star forming galaxy with  $z\! =\! 0.546$.

\noindent
{\bf b. Interpretation:}
The prompt SHB was produced by ICS of glory light as evident
from its spectral index $\Gamma\!\sim\!1$. The 
XRT light curve is well fit by the  SR tail of prompt 
flares taken over by the canonical SR afterglow emitted by a CB moving 
in a constant density ISM  of the host galaxy, as shown in
Fig.~\ref{F1}e. 
Its late afterglow with $\alpha\!=\!1.6\!\pm\! 0.05$ and $\Gamma\!=\!1.97 
\!\pm\! 0.13$ are consistent within errors with
the CB model prediction, $\alpha\!=\!\Gamma\!-\!1/2$.
The wiggling of the 
light curve around the smooth theoretical line can be due to density 
variations along its trajectory, as was
observed in afterglows of many LGRBs.

\noindent
{\bf SHB 051227.}\\
{\bf a. Observations:} This was a multi-peaked burst  
with a bright X-ray afterglow localized by Swift (Barbier et al.~2005).
It was initially thought
to be LGRB, since it has $T_{90}\!=\! 8.0\! \pm \!0.2$ s (Hullinger et 
al.~2005), 
but a spectral
analysis that showed a negligible spectral lag and 
a broad softer bump between
30 and 50 s were claimed to indicate an SHB (e.g. Kann et al.~2008).
An exceedingly faint optical afterglow was discovered 
within the XRT error circle (Malesani et al.~2005a,b)
with the VLT and also detected by Gemini.

\noindent
{\bf b. Interpretation:}
The prompt burst is consistent with ICS of glory light (
$\Gamma\!\sim\! 0.91\! \pm \! 0.28$). The XRT light curve is consistent 
with 
an SR tail of a prompt emission taken over by a canonical 
SR afterglow, as shown in Fig.~\ref{F1}f. The XRT does not shed light on 
whether this GRB was or was not an SHB. Due to a gap in the data 
between 400 s and 4000 s, the  theoretical shape and the values
of the SR afterglow parameters are uncertain.

\noindent
{\bf SHB(?) 060121.}\\ 
{\bf a. Observations:}
This intermediate duration GRB was discussed in detail in Donaghy et 
al.~2006, de Ugarte Postigo et al.~2006 and Levan et al.~2006.
It was localized by HETE-2. It consisted of two peaks with 
$T_{90}\!=\!1.60\! \pm\! 0.07$ s at high energies, which were followed by 
a faint soft emission for several hundreds of seconds.
It was also detected by Konus-Wind, Suzaku and RHESSI. 
Follow-up observations by Swift began 3 h
after the burst. Its XRT-light curve is shown in Fig.~\ref{F1}f.
Its  faint optical afterglow was discovered  by Malesani et al.~(2006), 
which led to the discovery of its extremely faint host galaxy (Levan et 
al. 2006). The burst outshined its host galaxy (by a 
factor $\!>\!100$). A photometric redshift for this event placed it
at a most probable redshift of $z\!=\! 4.6$, with a less probable 
$z\! =\! 1.7$. In either case, GRB 060121 could be the farthermost 
short GRB detected to date with an isotropic-equivalent 
energy release in gamma rays comparable to that of LGRBs.

\begin{figure}[]
\centering
\vspace{-1cm}
\vbox{
\hbox{
\epsfig{file=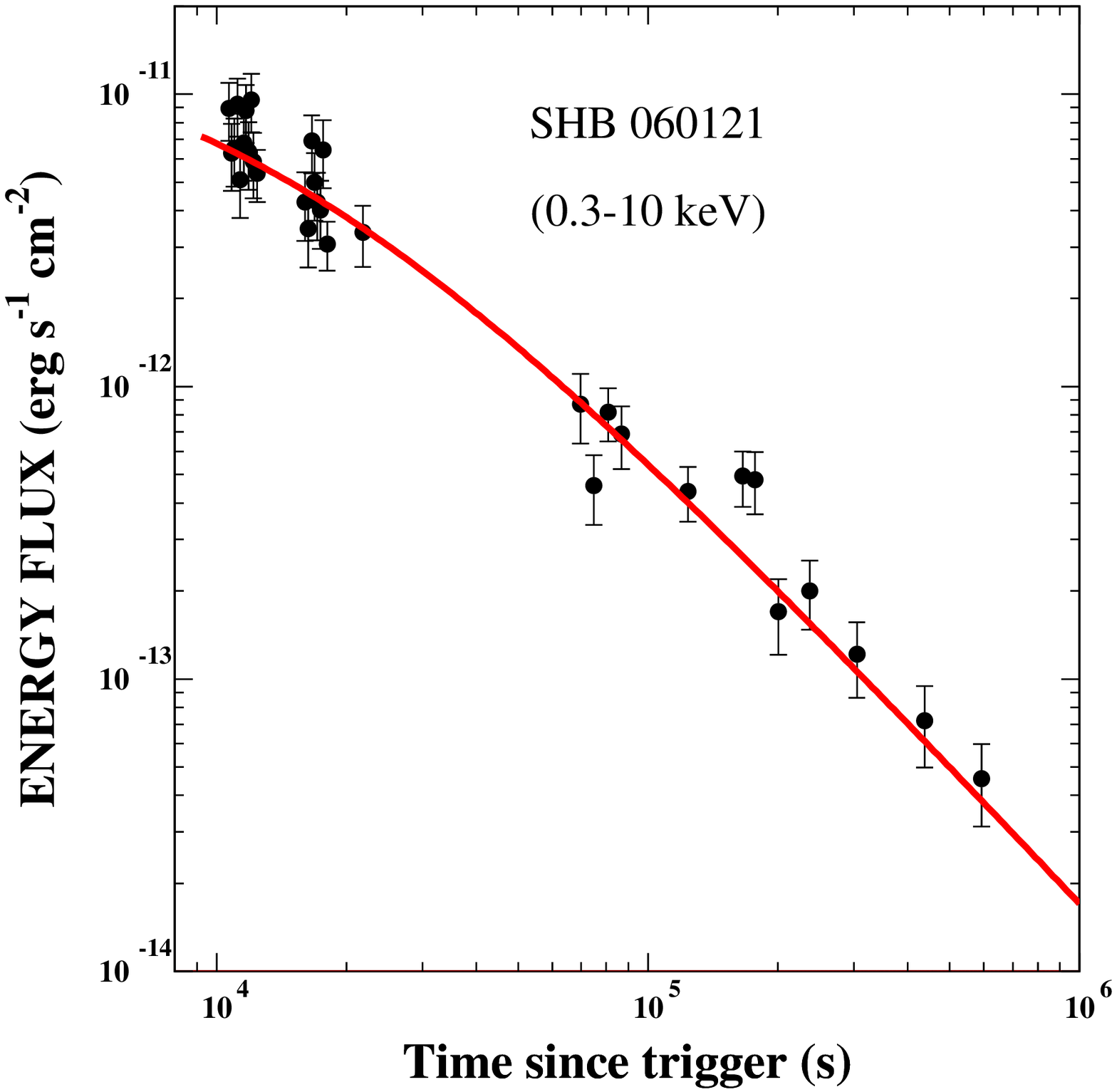,width=8.0cm,height=6.0cm}
\epsfig{file=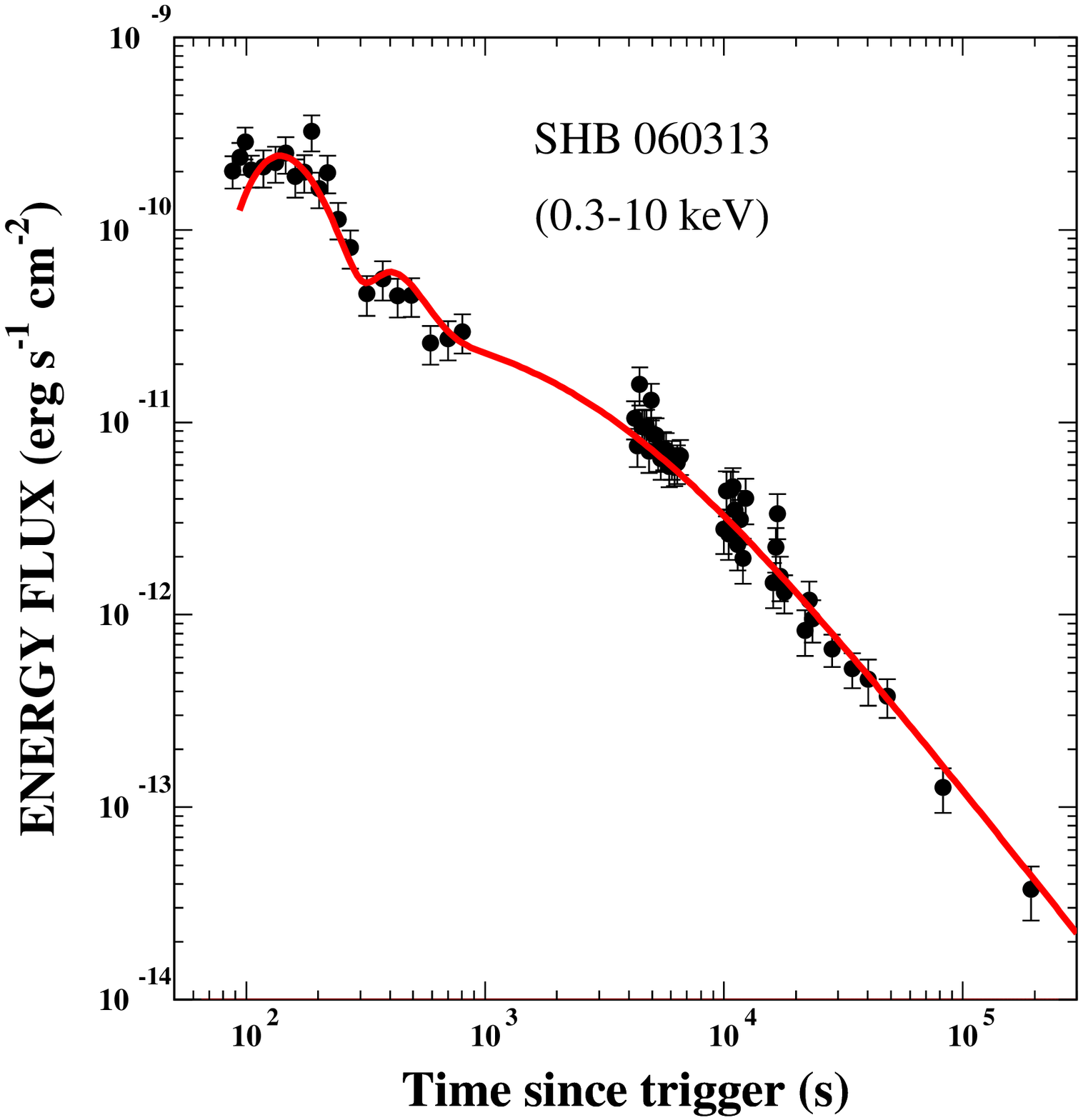,width=8.0cm,height=6.0cm}
}}
\vbox{
\hbox{
\epsfig{file=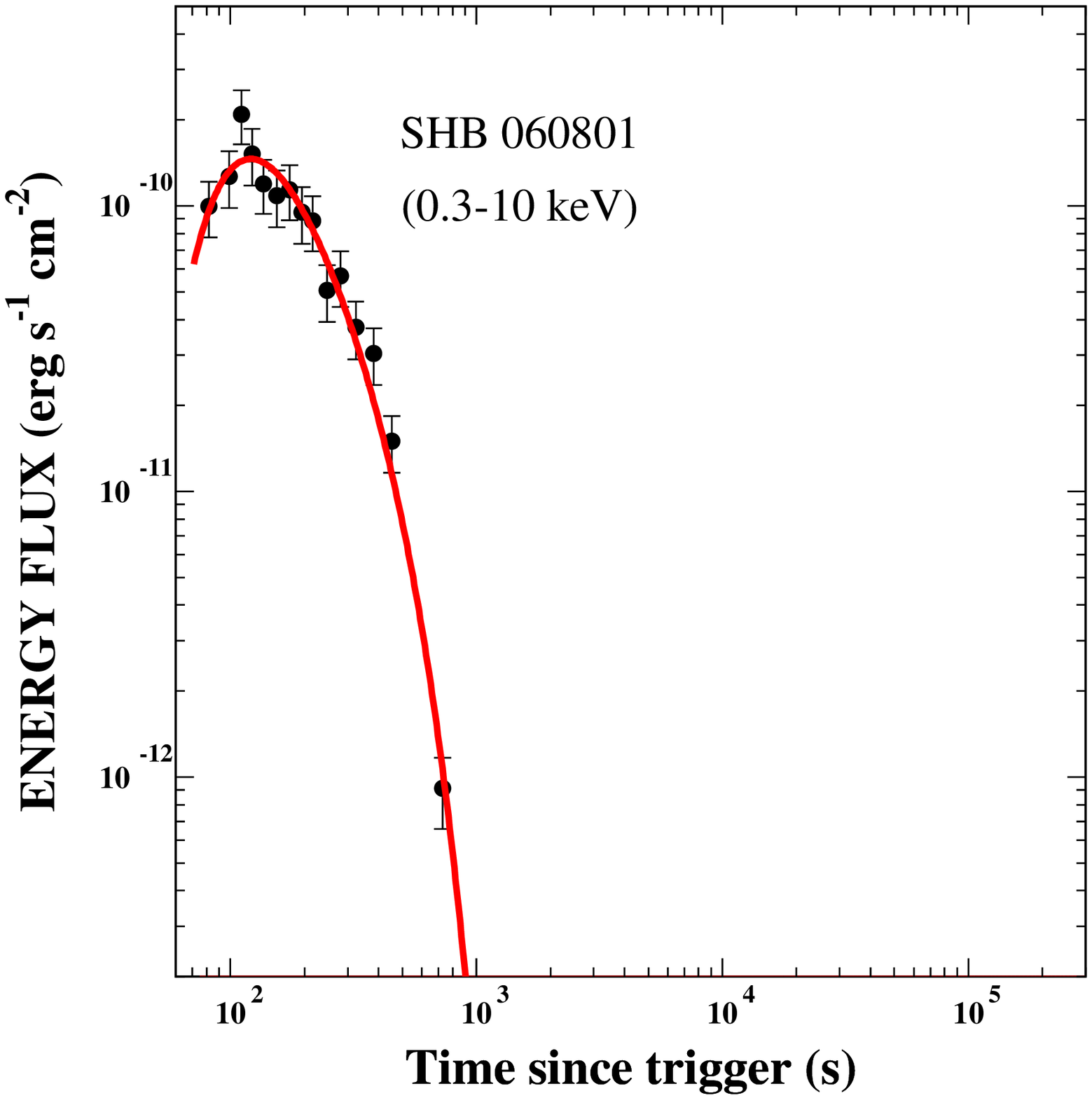,width=8.0cm,height=6.0cm}
\epsfig{file=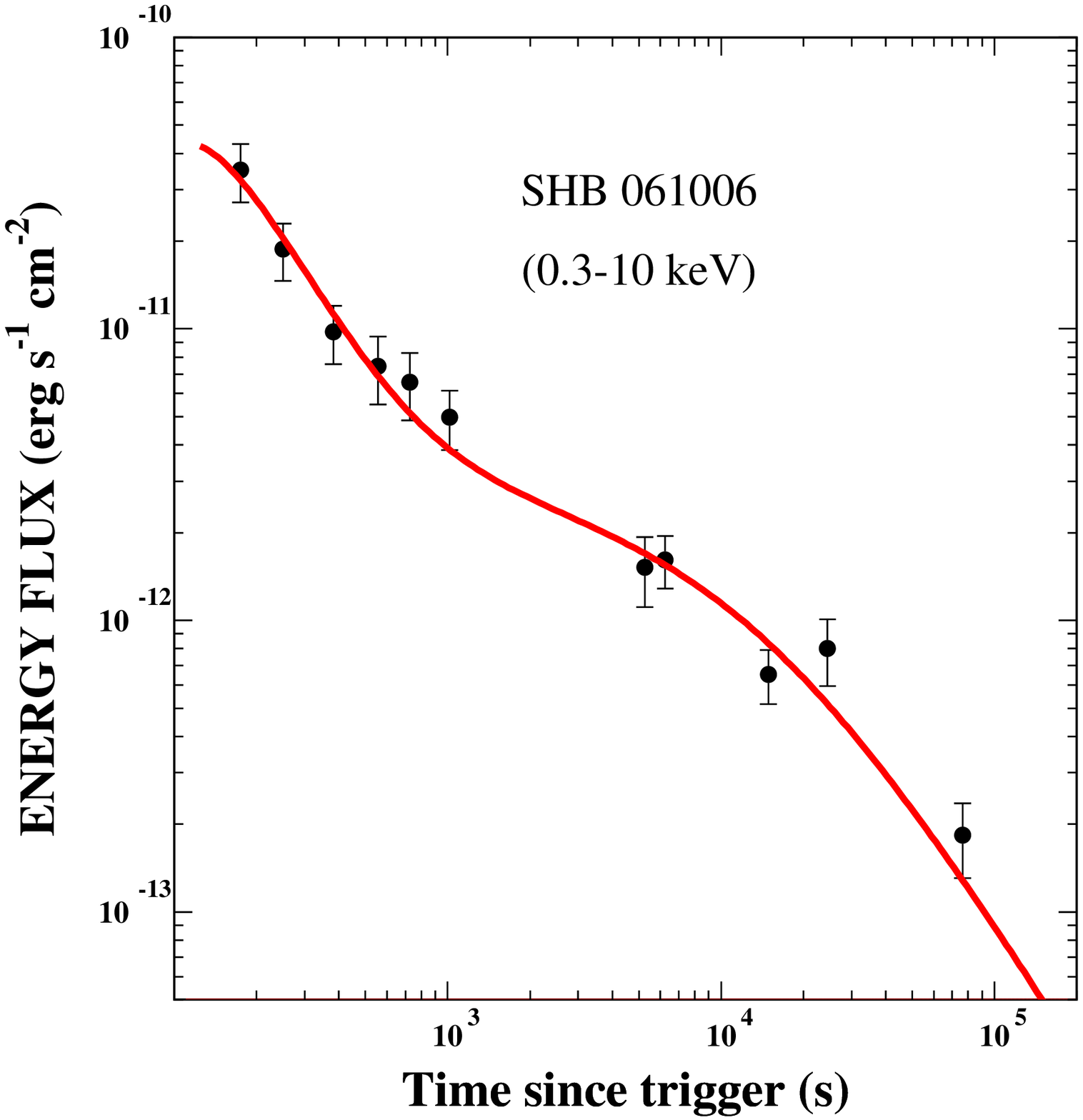,width=8.0cm,height=6.0cm}
}}
\vbox{
\hbox{
\epsfig{file=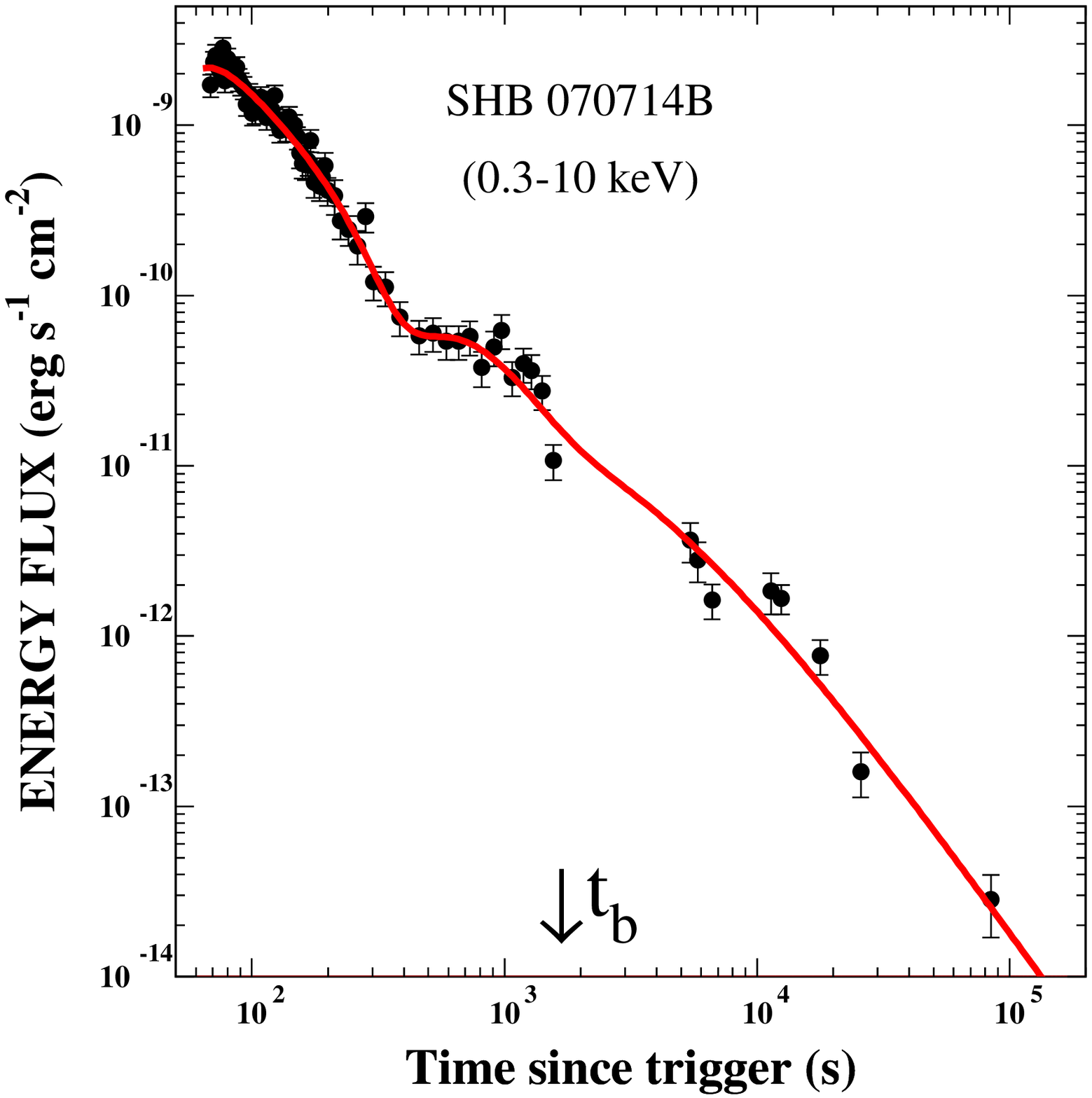,width=8.0cm,height=6.0cm}
\epsfig{file=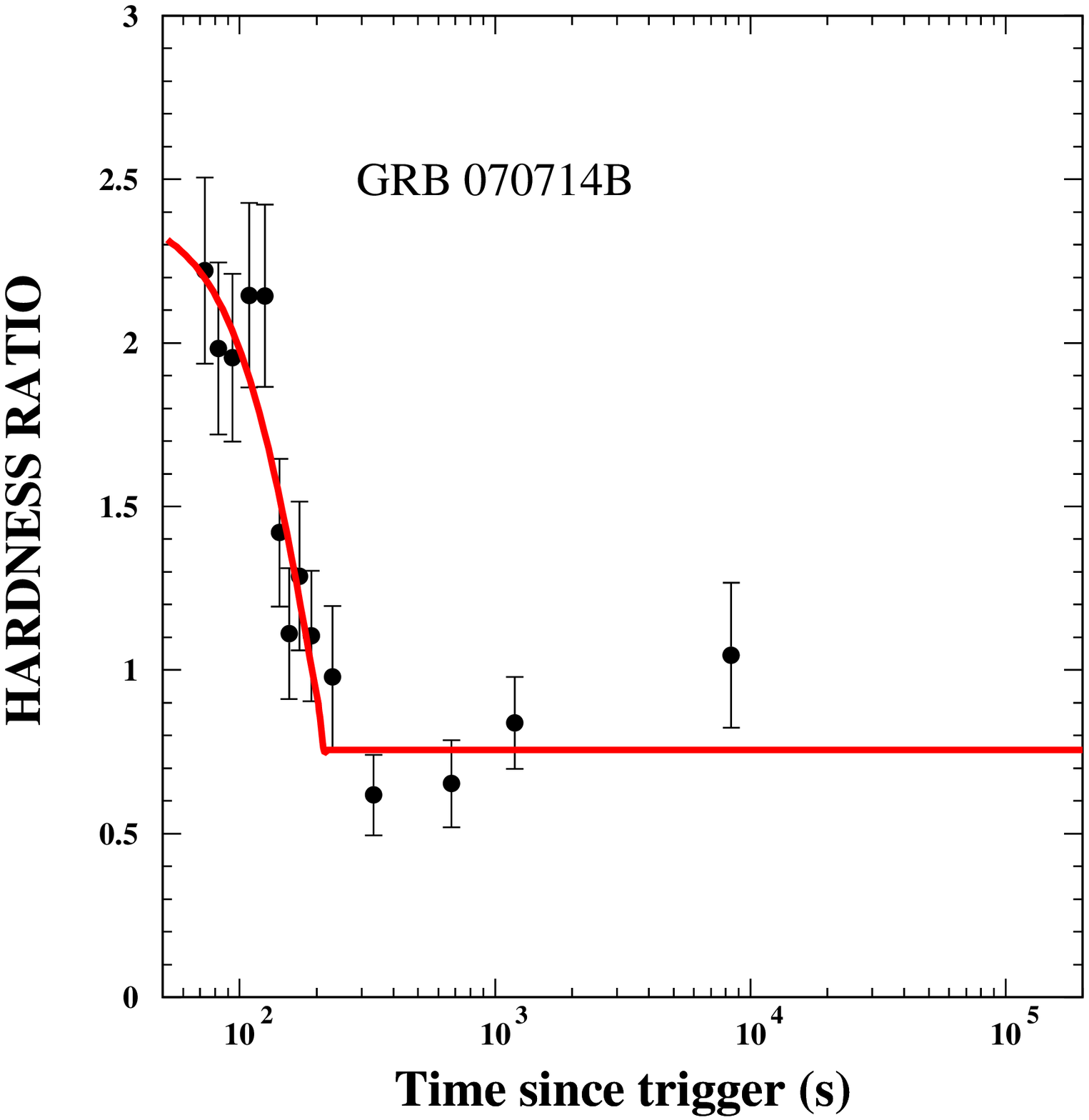,width=8.0cm,height=6.0cm}
}}
\caption{
Comparisons between SWIFT XRT observations of SHBs
(Evans et al.~2007) and the CB model predictions.
{\bf Top left (a):} The light curve of SHB 060121.
{\bf Top right (b):} The light curve of SHB GRB 060313.
{\bf Middle left (c):} The light curve of SHB 060801.
{\bf Middle right (d):} The light curve of SHB 061006.
{\bf Bottom left (e):} The light curve of SHB 070714b
{\bf Bottom right (f):} The hardness ratio of 070714b.
}
\label{F2}
\end{figure}

\noindent
{\bf b. Interpretation:}
Its intermediate  duration and 
its unusual large redshift $Z\!=\!4.7$,  relatively small $E_p\!\approx 
\! 120$ keV, large equivalent isotropic gamma ray energy,  
$E_{iso}\!\approx\! 2.2 \times 10^{53}$ erg 
(or $E_{iso}\!\approx\! 4.3 \times 10^{52}$ erg for $z\!=\!1.7$), vicinity
($\!\sim\! ~2$ kpc ) to the center of the host galaxy,  
 and its bright optical afterglow, suggest that GRB 
060121 may have been an ordinary LGRB, and was wrongly classified as SHB. 
In Fig.~\ref{F2}a we show the CB model fit to its late AG,
which, although satisfactory, does not shed light on its true identity.

\noindent
{\bf SHB 060313}.\\
{\bf a. Observations}:
This unusual burst was reported and discussed in detail in Roming et 
al.~2006. It was detected by Swift, Konus-Wind and INTEGRAL.
It was a very  intense burst, with many short spikes each with
FWHM smaller than 20 ms, with $T_{90}\! =\! 0.7\! \pm\! 0.1$ s.
It had a very unusual average initial photon index, 
$\Gamma\!=\!-0.33(-0.29,+0.25)$ below the spectral peak
during the initial 0.192 s of the burst (Golenetskii et al.~2006a).
It had  the highest fluence and the highest observed peak
energy of all SHBs observed by Swift. No extended soft emission was
detected.
Swift XRT detected its X-ray afterglow 79 s after the BAT trigger
and measured its light curve  until 200 ks (Fig.~\ref{F2}b) with a
late afterglow photon index of $\Gamma\!=\!1.96\! \pm\! 0.09\, .$
The XRT light curve shows flaring activity  
superimposed on a canonical light curve during the  first 1000 s after 
burst.
The optical afterglow of SHB 060121 was detected and followed by the Swift 
UVOT and by the VLT following its localization by Swift UVOT. A very 
faint host galaxy was detected at the afterglow position
by Berger et al.~(2006a). 

\noindent
{\bf b. Interpretation:}
The unusual early light curve and spectrum of SHB 060313, suggest an 
unusual 
progenitor and environment of this unusual SHB. In the framework 
of the CB model, the large number of short pulses requires 
a progenitor which fires many small
CBs  like a shot gun (Plaga, private communication) or
a machine gun, rather than a cannon. 
The unusual hard spectrum could be produced by inverse Compton 
scattering of self absorbed radiation produced inside the CBs. 
In contrast to the unusual prompt emission, the 
afterglow is not unusual, and can be well explained as SR 
radiation from a CB  propagating in a constant density 
ISM of the host galaxy, as shown in Fig.~{F2}b.

\noindent
{\bf SHB 060801}.\\
{\bf a. Observations:}
This SHB was detected and localized by Swift  (Racusin et al.~2006a)
and was also detected by
Suzaku. It had $T_{90}\!=\! 0.5\!\pm\! 0.1$ s. It consisted of
two pulses which peaked 60 ms and $\sim$ 100 ms after their
beginning (Sato et al.~2006). No extended soft emission was detected. 
The XRT began taking data 63 s after the BAT trigger (Racusin et al.~2006b).
The 0.3-10 keV X-ray light curve shows a plateau from 73 s until 115 s 
after the BAT trigger, which turns into an exponential decay.
The X-ray emission was not detected after 1000 s. 
Follow up optical observations did not detect an optical afterglow. 
A single galaxy lying within the XRT error box at redshift 
$z\!=\!1.1304$  was suggested as its host galaxy, making 
SHB 060801 the most distant SHB with a secured redshift.

\noindent
{\bf b. Interpretation:}
The prompt emission pulses are well explained by ICS of   
a thin bremsstrahlung glory of two CBs. The extended soft emission at 
redshift 1.13 was too dim to be detected by BAT, but its ending and 
decay probably are the emission detected and followed with the XRT.
In the CB model it is well described by  ICS of the GC's light 
when the CBs leave its dense core and move into the surrounding 
ISM,  as shown in Fig.~\ref{F2}c.

\noindent
{\bf SHB 061006.}\\
{\bf a. Observations:}
The Swift observations of this burst are reported in 
Schady et al.~2006. This burst 
began with an intense double spike which lasted 0.5 s. The SHB was also 
detected by RHESSI, KONUS-WIND and Suzaku (Hurley, et al.~2006). It was 
followed by an extended soft emission with $T_{90}\!=\! 130\!\pm\!10$ s. 
The 
XRT began follow-up observations at 143 s after burst. The XRT light curve 
(Fig.~\ref{F2}d) initially decayed with a slope of $\alpha\!=\!2.3\!\pm\! 
0.3$, followed by a shallow slope beginning at 300 s with a spectral index 
$\Gamma\!=\!1.7\!\pm\!0.3$. VLT observations (Malesani et al.~2006b) 
found a  faint source that was subsequently found to fade, revealing a 
starforming  host galaxy (Malesani et al.~2006c, Berger et al.~2006b) at 
redshift $z\!=\! 0.4377\,.$ The burst location is about $8\,kpc$ from 
the galactic center.

\noindent
{\bf b. Interpretation:}
In the CB model, the prompt pulses can be explained by ICS of glory light
of the SHB progenitor. The extended 
soft emission could be produced by synchrotron radiation from
the CBs while they were crossing the superstar cluster where the SHB took 
place. As shown in Fig.~\ref{F2}d, the initial decay of the XRT light 
curve can be explained 
by SR emission from the CBs while it was crossing the SSC wind, 
whereas the shallow decaying AG is the afterglow emitted when the CBs 
propagate in the ISM surrounding the SSC.

\noindent {\bf SHB 070714B.}\\ 
{\bf a. Observations:} The Swift observations of this SHB are detailed in 
Racusin et al.~2007. The SHB was a very bright multi-spiked event lasting 
about three seconds which was followed by soft extended emission that 
lasted $64\!\pm\! 5$ s. Swift/XRT began follow-up observations 61 sec 
after 
trigger. A joint Swift - Suzaku spectral analysis yielded a hard 
spectrum and a high peak energy (Ohno et al.~2007). The photon spectral 
index integrated over the SHB was $\Gamma=0.99\!\pm\! 0.08$. The 0.3-10 
keV light curve (Fig.~\ref{F2}) showed the canonical behaviour observed  
in LGRBs, 
namely, a steep decay with super-imposed  flares  taken 
over around 400 s by an afterglow that gradually bends over into an 
asymptotic power-law decay with $\alpha\!\sim\! 1.6$. 
The fast decay of the soft extended emission 
was accompanied by a rapid spectral softening (Fig.~\ref{F2}f)
until the plateau phase took over. The photon spectral index
during the AG phase  was $\Gamma\!=\!1.95 \pm 0.15$  (Zhang et al.~2007).
An optical afterglow was discovered 10 minutes after the GRB by the 
Liverpool Telescope (Melandri et al.~2007), which led to the discovery 
of a host galaxy  at a redshift $z\!=0.92$ 
(Graham et al.~2008). 

\noindent
{\bf b. Interpretation:}
The initial fast decay and rapid spectral evolution of the 
early time X-ray light curve of SHB 070714B,  are well reproduced 
by the CB model (Figs.~\ref{F2}e,f. The data beyond 400 s
show a considerable flaring activity. This makes quite uncertain  
the CB model 
best fitted parameters of the smooth SR afterglow component, reported 
in Table II.

\noindent
{\bf SHB 070724A.}\\
{\bf a. Observations:} This short burst was 
a faint single-spiked GRB with ${\rm T_{90} = 0.40\!\pm\! 0.04}$ s 
localized by Swift with no detected ESEC (Ziaeepour et al.~2007). 
Swift XRT began its
observations of the burst X-ray afterglow 72.1 s after trigger and 
followed its rapid decay and two 
superimposed  X-ray flares peaking around 127 s 
and 200 s. The fast decay of the AG turned into 
a shallow decay around 400s 
which steepened around 40 ks.
A faint source at redshift  $z\! =\! 0.457$ within the XRT error circle
was suggested as a possible star forming host galaxy.
Deep imaging and image subtraction did not 
reveal an optical afterglow. 
An analysis of the BAT data yielded  
$E_p\!\sim\! 41$  keV and a photon index of  $\Gamma= 2.2$. 

\begin{figure}[]
\centering
\vspace{-1cm}
\vbox{
\hbox{
\epsfig{file=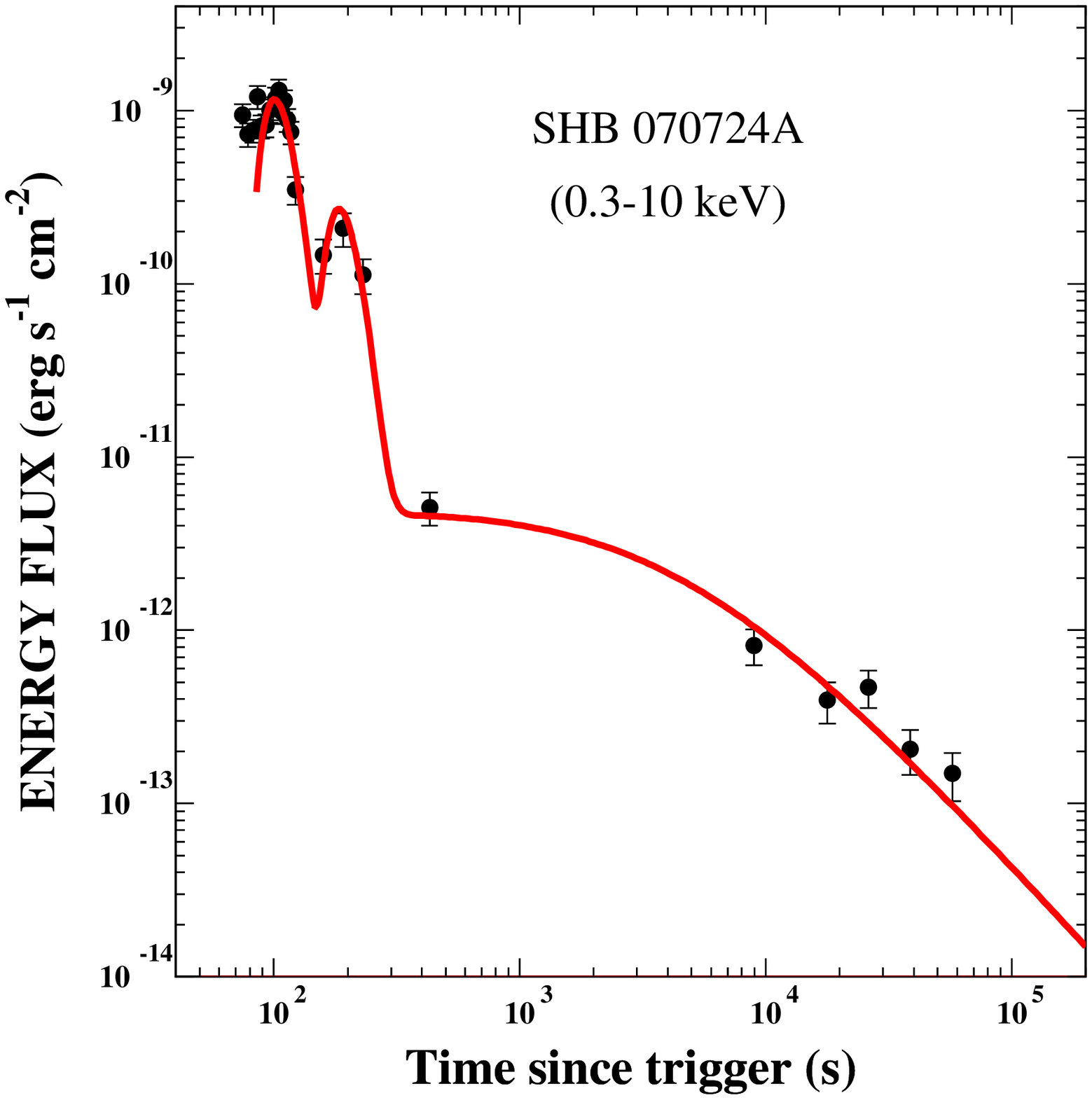,width=8cm,height=6.0cm}
\epsfig{file=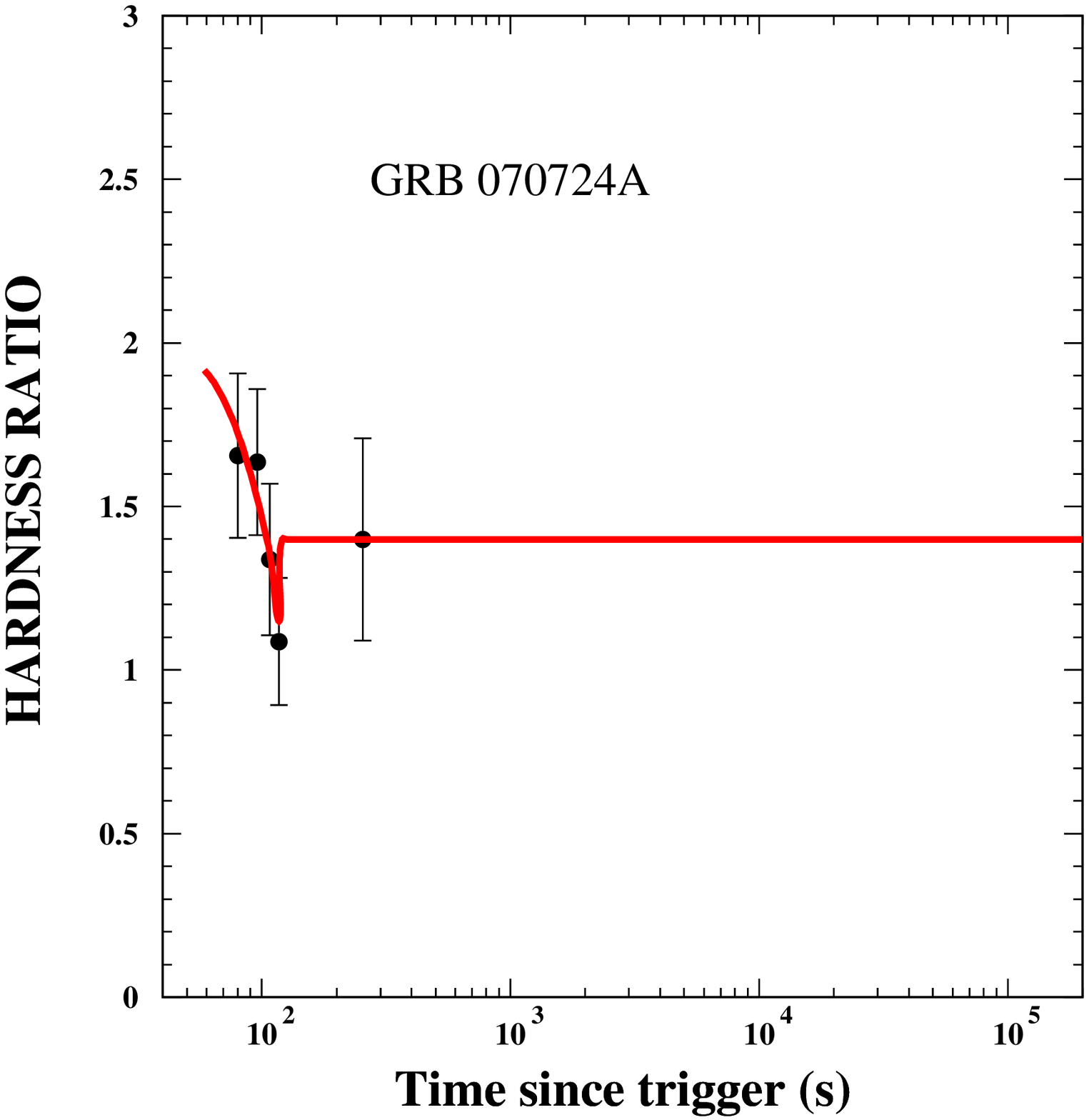,width=8cm,height=6.0cm}
}}

\vbox{
\hbox{
\epsfig{file=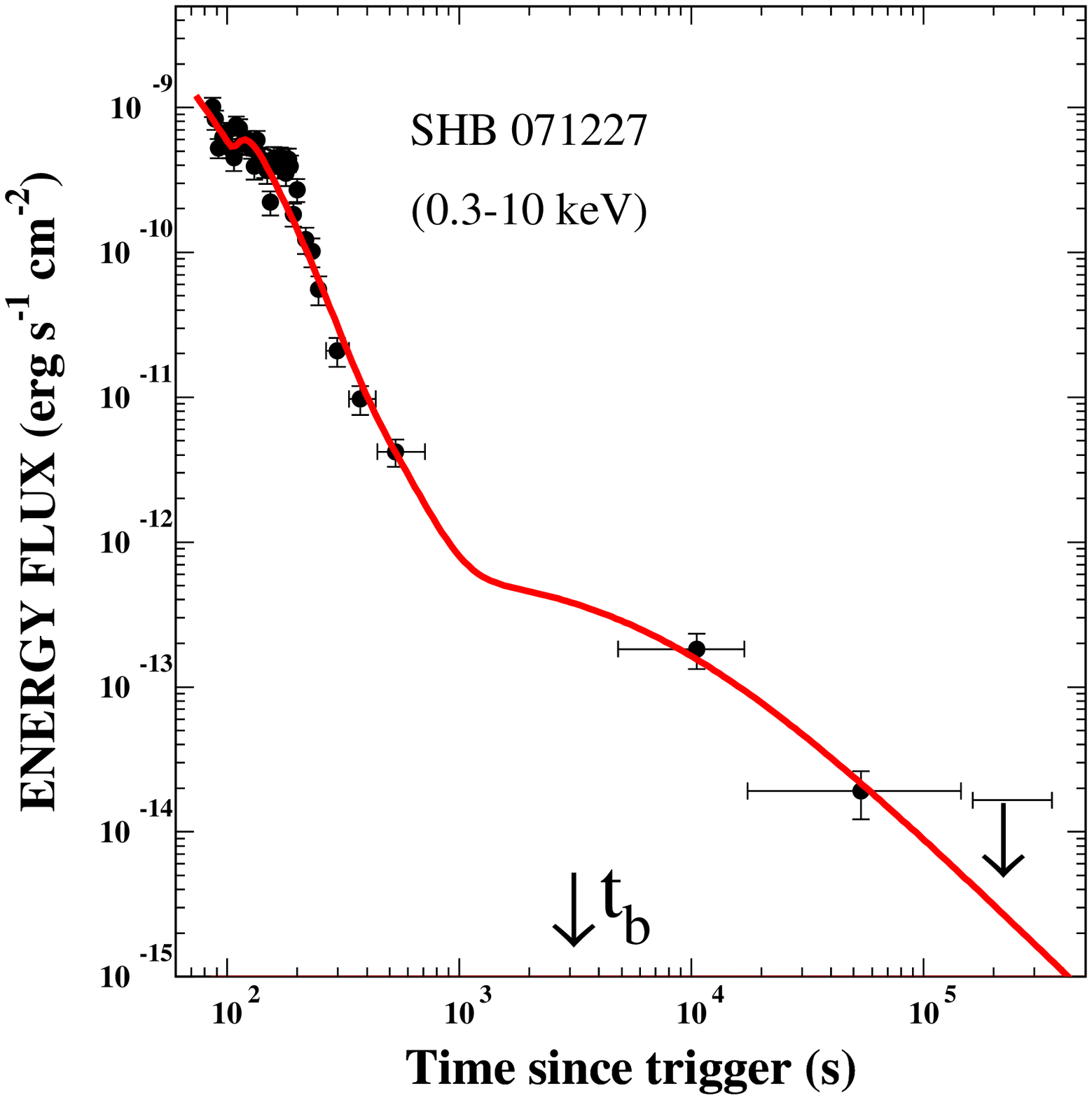,width=8.0cm,height=6.0cm}
\epsfig{file=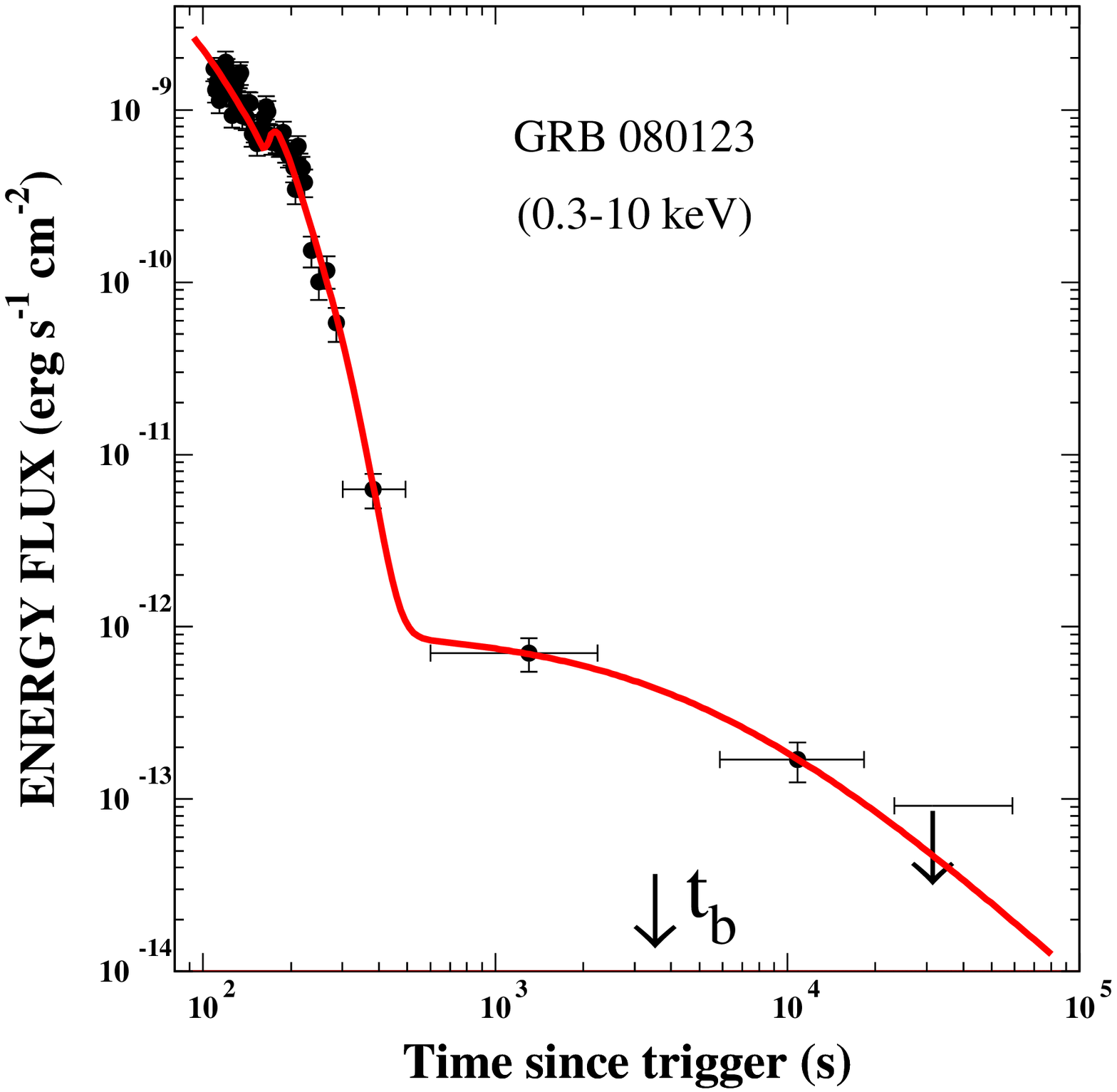,width=8.0cm,,height=6.0cm}
}}

\vbox{
\hbox{
\epsfig{file=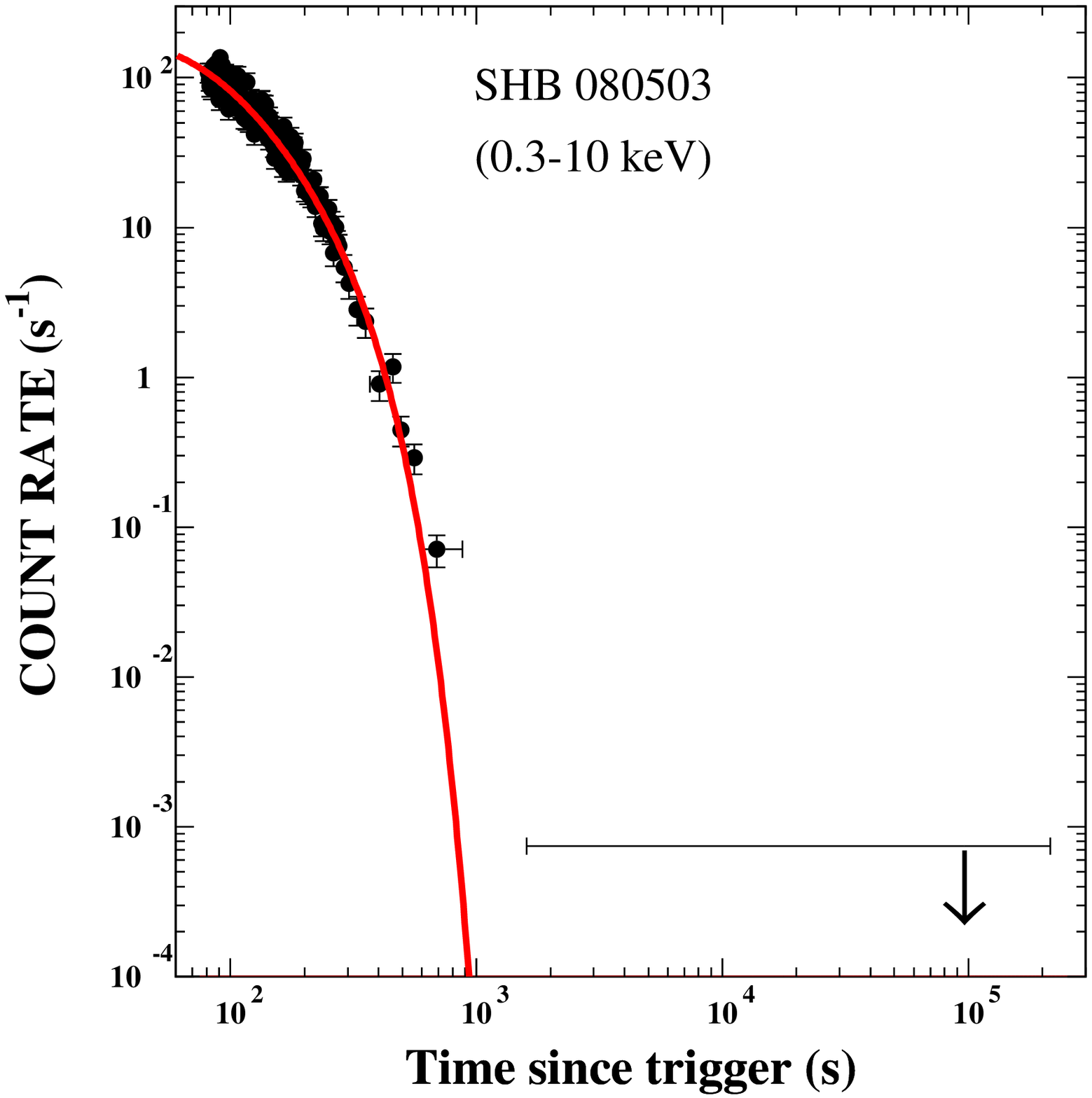,width=8cm,height=6.0cm}
\epsfig{file=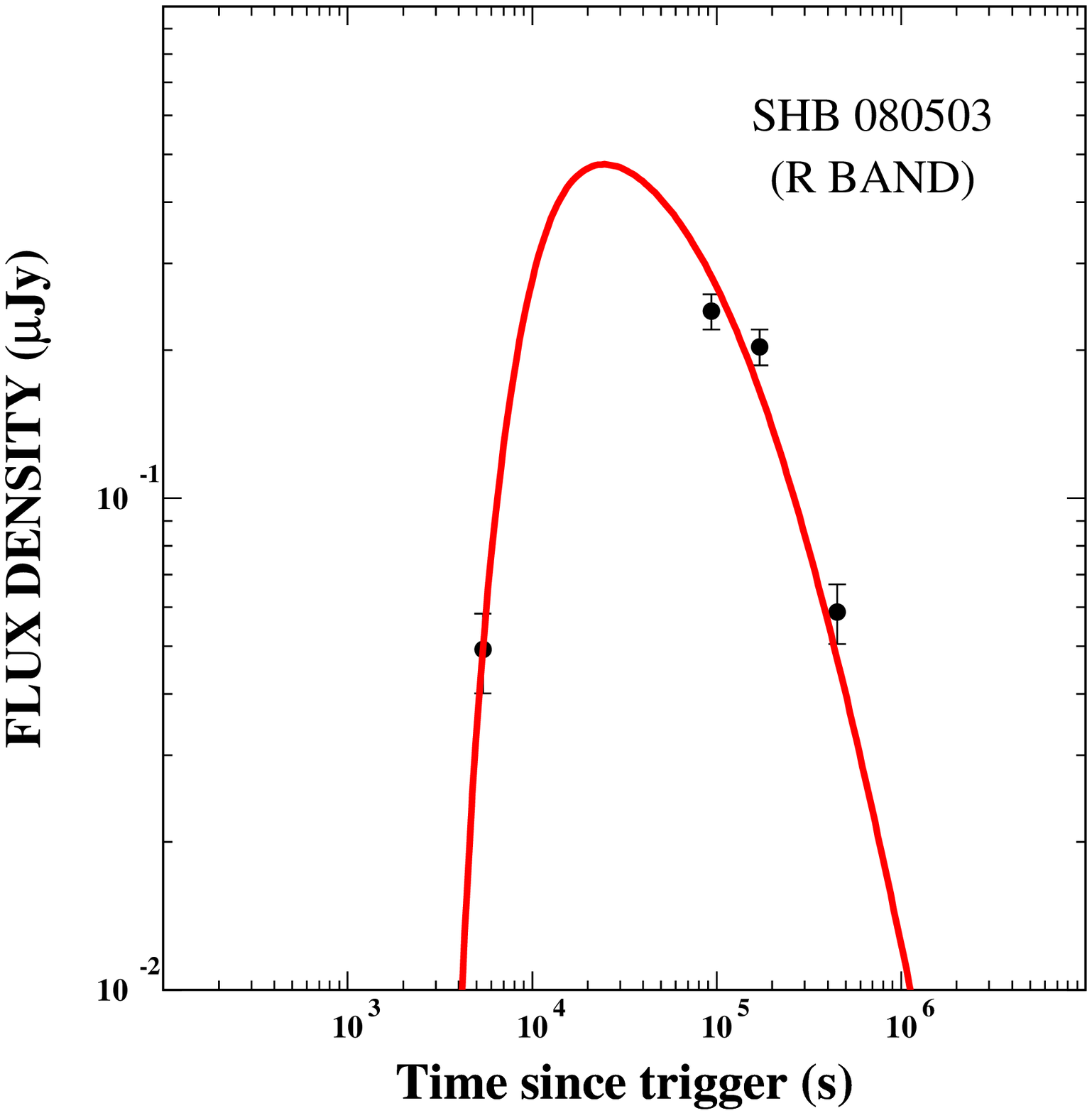,width=8.0cm,height=6.0cm}

}}
\caption{
Comparisons between observations of SHBs and
CB model predictions.
{\bf Top left (a):} The XRT light curve of SHB 070724A.
{\bf Top right (b):} The hardness ratio of SHB 070724A.
{\bf Middle left (c):} The XRT light curve of SHB 071227.
{\bf Middle right (d):} The XRT light curve of SHB 080123.
{\bf Bottom left(a):} The XRT light curve of SHB 080503.
{\bf Bottom right (b):} The R-band light curve of SHB 080503.
}
\label{F3}
\end{figure}

\noindent
{\bf b. Interpretation:}
The CB model fit to the XRT light curve is shown in 
Fig.~\ref{F3}a. The light curve is well fitted by two ICS 
pulses taken over by an SR afterglow from a CB moving in a constant 
low-density  ISM. The relatively large value,  $\gamma\, \theta$ 
indicates nearly a `far off axis' SHB consistent with its relatively 
small $E_{iso}$ and $E_p$ and a large pulse width compared to those 
of ordinary SHBs. 
  
\noindent
{\bf SHB 071227A.}\\
{\bf a. Observations:} This short burst was detected and localized by 
Swift (Sakamoto et al.~2007, Sato et al.~2007). It was 
also observed by Suzaku and Konus-Wind. 
The SHB was a multi-peaked structure
with a duration of about 1.5 seconds and a time averaged 
power-law spectrum with a photon spectral index $\Gamma\!=\!1.2\!\pm\! 
0.2$.
The Swift XRT began observing GRB 071227 79.5 seconds after the BAT 
trigger. The XRT light curve shows the canonical behaviour: soft flares 
superposed on a decaying light curve which changes to 
a fast decay
with a spectral softening, until it  is taken over by a plateau/shallow
decay  around  1000 seconds. 

\noindent
{\bf b. Interpretation}
The CB model fit to the XRT light curve is shown in
Fig.~\ref{F3}c. The light curve and spectral evolution 
are well fitted by ICS of a GC light of a CB moving away 
from the GC into a constant low-density  ISM
where the AG is dominated by  SR emission.
However, the data after the fast decay phase are
scarce and  constrain only weakly the CB model parameters 
of the SR afterglow.

\noindent
{\bf SHB 080123.}\\
{\bf a. Observations:} This short burst was detected and localized by 
Swift. The Swift observations of this burst are reported in detail 
in Ukwatta et al.~2008 and Copete et al.~2008. 
Its light curve  shows
two well-separated peaks. The 
first started at 0.3 s
and is no wider than 64 msec. The second peak started at 0.6 s 
with a FRED-like shape and
a duration of 256 ms. Its soft extended emission lasted 120 s
after the prompt hard emission of the short GRB
and  included soft flares (Copete et al.~2008).  
The XRT observations started 108 s after the BAT trigger.
The XRT light curve (Fig.~\ref{F3}d)
shows the canonical behaviour: a fast decay 
with a rapid spectral softening, which was taken over by a plateau/shallow
decay  around 500 s with a typical photon index of 
$\Gamma\!=\! 2.1 \!\pm\! 0.2$.

\noindent
{\bf b. Interpretation:}
The fast decay of the XRT early time light curve 
and its rapid spectral softening (not shown here) 
are well fit by ICS of light from a GC/SSC.
It is  being taken over  around 500 s
by a plateau/shallow decay 
with a typical photon index of $\Gamma\!=\! 2.1\! \pm
\!0.2$ which is well fit by SR emission from a CB deceleration in a 
a medium of low constant density (Fig.~\ref{F3}d).
The scarce data on the late afterglow constrain only weakly the 
CB model parameters of the SR afterglow. 

\noindent
{\bf SHB 080503.}\\
{\bf a. Observations:} This short burst was detected and localized by 
Swift. The Swift observations of this burst are reported in detail in Mao 
et al.~2008. Its light curve (Fig.~\ref{F3}e) shows an initial spike 
starting at 0.1 s after the BAT trigger with a fast rise to a peak at 0.2 
s, then, a roughly exponential decay down to background at 0.7 s. It had 
an extended soft emission which started at about 10 s, rose with two peaks 
at 26 s and 37 s, and then fell to background levels at 220 s. The Swift 
XRT began observing SHB 080503  81 s after the BAT trigger. The XRT 
light curve showed an exponential decay with a gradual spectral 
softening from 
$\Gamma\!\sim 1$ at the beginning of the XRT observations to $\Gamma\!\sim 
3$ around 500 s. The X-ray AG (Fig~\ref{F3}e) decreased below the Swift 
detection sensitivity around 1000 s after burst, but was detected later 
between 4.29-4.66 days after burst by Chandra (Butler et al.~2008). A 
rising optical afterglow was detected on the second night after trigger by 
Perley et al.~(2008a) with Gemini-North. Continuous monitoring of this 
optical counterpart on consecutive nights (Bloom et al.~2008,
Perley et al.~2008b) found no 
further re-brightening and its fading (Fig.~\ref{F3}f). The OT was not 
detected with HST 9.2 days after the BAT trigger (Perley et al.~2008c).

\noindent
{\bf b. Interpretation:}
The exponential 
decay of the  XRT light curve at the end of the 
ESEC is well described by ICS of 
a CB emerging from the dense stellar core of a superstar cluster of 
a typical core radius  $r_c\sim 1$ pc, as shown in Fig.~\ref{F3}e.
This is supported by the rapid spectral softening/decreasing
hardness ratio during this decay. The decay is stretched over
$t > 900$ s, probably because of a relatively large redshift 
and a very low surrounding density which results in a very faint
X-ray synchrotron afterglow  and  a delayed  take-over time.
The low extra cluster density yields a prolonged  expansion time
of the CB, $t_{exp}\!\approx\!7553 $, in Eq.~(\ref{SRSSC})
 and an initially rising  SR afterglow 
(Fig.~\ref{F3}f). The 
asymptotic decay of the optical AG is well described by $F_\nu\propto 
t^{-(1+\beta)}$ (Fig.~\ref{F3}f)
with $ \beta\!=\!\beta_X\! \approx\! 1.1$,
typical of a SR emitted by a CB moving 
in the wind of a young superstar cluster (a proto globular cluster).
A superstar cluster environment of the SHB is supported by the 
general properties of its ESEC.

\noindent
{\bf GRB 060614.}\\ 
{\bf a. Observations:}
This puzzling GRB was discussed in detail 
in Gehrels et al.~2006, 
Gal-Yam et al.~2006 Cobb et al. 2006, and Mangano et al.~2008. 
Swift-BAT triggered on GRB 060614 on
2006 June 14 at 12:43:48 UT. The  BAT light curve showed a 
5 s series of short hard
peaks followed by a fainter, softer and highly
variable extended prompt emission with $T_{90}\!\sim\! 102$ s.
Konus-Wind was also triggered by GRB 060614, about 4 s after the BAT
trigger. For the initial group of short hard peaks it measured  
$E_p\!\sim\! 300$ keV. 
The XRT and UVO aboard Swift started observation 97 s after the 
BAT trigger which continued up to 51 days after burst 
The X-ray light curve showed the canonical behaviour observed 
in many LGRBs and in practically all SHBs, i.e. initial fast decay with 
a rapid spectral softening 
which is taken  over by a plateau that later bends into an asymptotic 
power-law 
decay. The AG bend/break started around 30 ks and the asymptotic 
temporal decay had a power-law index $\alpha\!=\!2.1\!\pm\! 0.07$. 
The burst was located at the outskirts of  a faint starforming 
galaxy at redshift $z=0.125$. Very deep searches (Gal-Yam et al.~2006  
Mangano et al.~2008) did not discover an associated SN.

\noindent
{\bf b. Interpretation:}
The initial short peaks of GRB 060614, 
the lack of an associate SN akin to those 
discovered in  ordinary LGRBs, the strict limits on  lag-times 
in its early short pulses, its extended soft emission
(well fit by a power-law spectrum with a photon index $\Gamma=2.13\pm 
0.05$) and its low isotropic energy ($(2.5\pm 0.7)\times 10^{51}$ erg;
Mangano et al.~2008) are typical of SHBs.
On the other hand, the long $T_{90}\!\sim\! 5$ s of its initial complex 
of short pulses, its $(1+z)\,E_p\!\sim\! 440\,(-281,+923)$ keV, and its 
location 
inside a starforming galaxy are typical of LGRBs.
Various solutions of these apparent contradicting pieces of evidence were 
suggested. Gal-Yam et al.~2006 and Gehrels et al.~2006 suggested that
GRB 060614 (and GRB 060505) belongs to a new class of long GRBs. 
Cobb et al.~(2006) suggested that its location in a the nearby host galaxy 
could be a chance coincidence. Dado et al.~(2008) suggested 
that GRB 060614 could be produced by an extremely 
faint core collapse supernova or by a ``failed supernova'' 
- the original collapsar model (Woosley 1993). 

Our CB-model analysis indicates that
GRB 060614 probably was a very hard and energetic SHB in a very bright SSC 
which was viewed far off axis ($\gamma\, \theta \ ~ 3.08$).
Its relatively large viewing angle yielded a relatively small Doppler 
factor 
$\delta$ which stretched the observed durations of its prompt emission
pulses and its extended soft emission component 
relative to those of ordinary SHBs,  and decreased 
their peak emission energy, equivalent isotropic energy and peak 
luminosity. 

\begin{figure}[]
\centering
\vspace{-1cm}
\vbox{
\hbox{
\epsfig{file=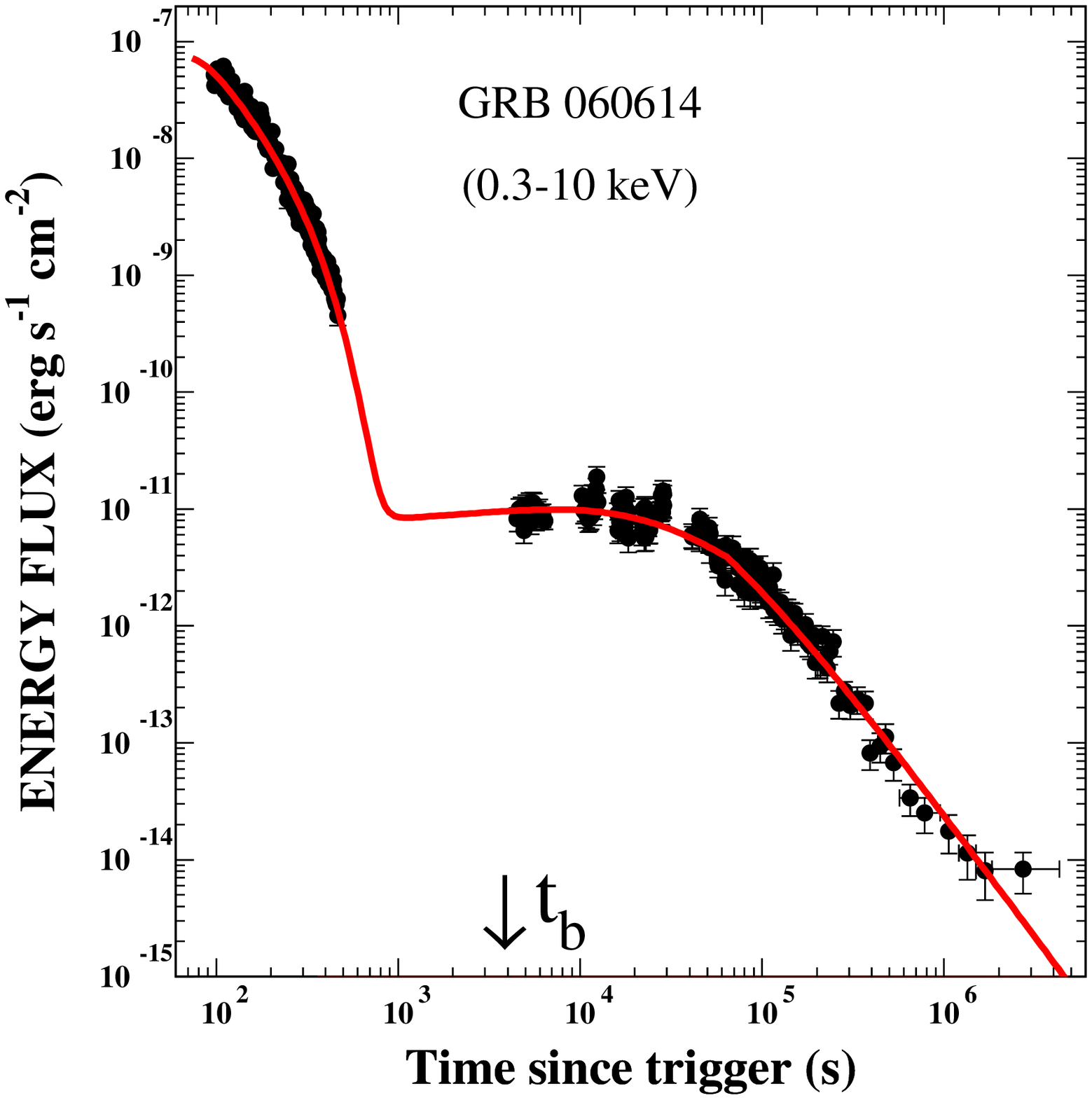,width=8.0cm,height=6.0cm}
\epsfig{file=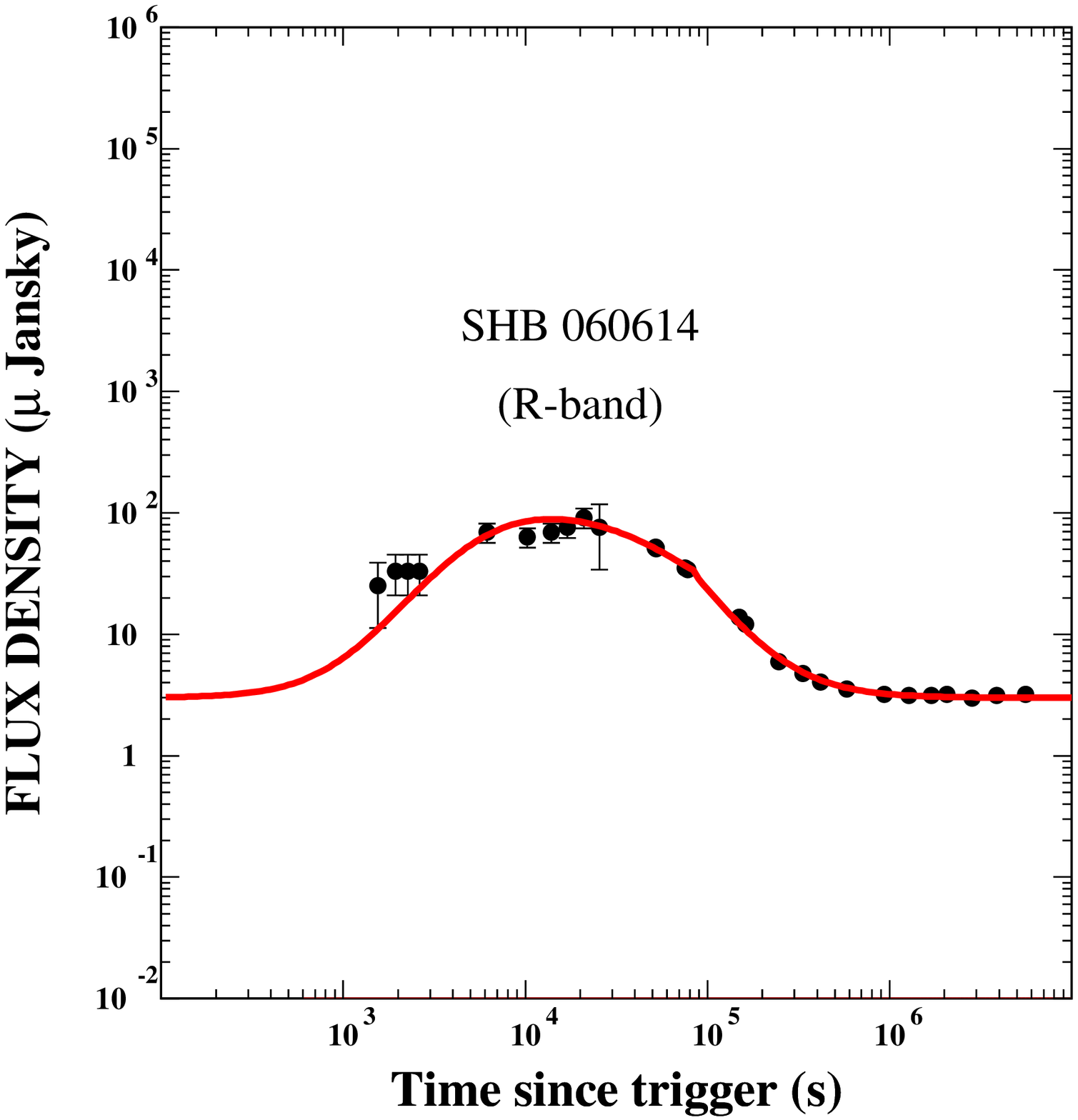,width=8.0cm,height=6.0cm }
}}
\vbox{
\hbox{
\epsfig{file=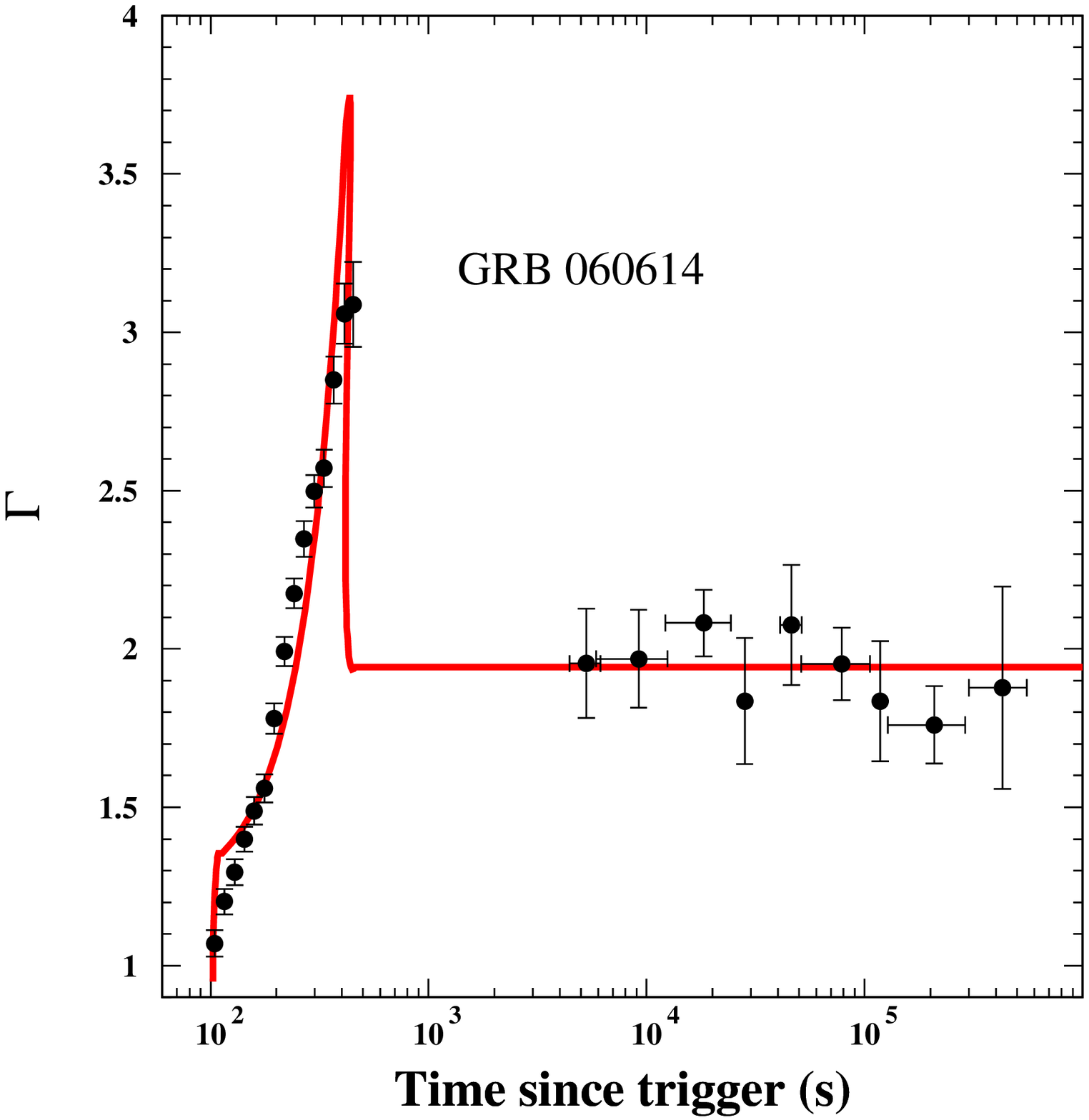,width=8.0cm,height=6.0cm}
\epsfig{file=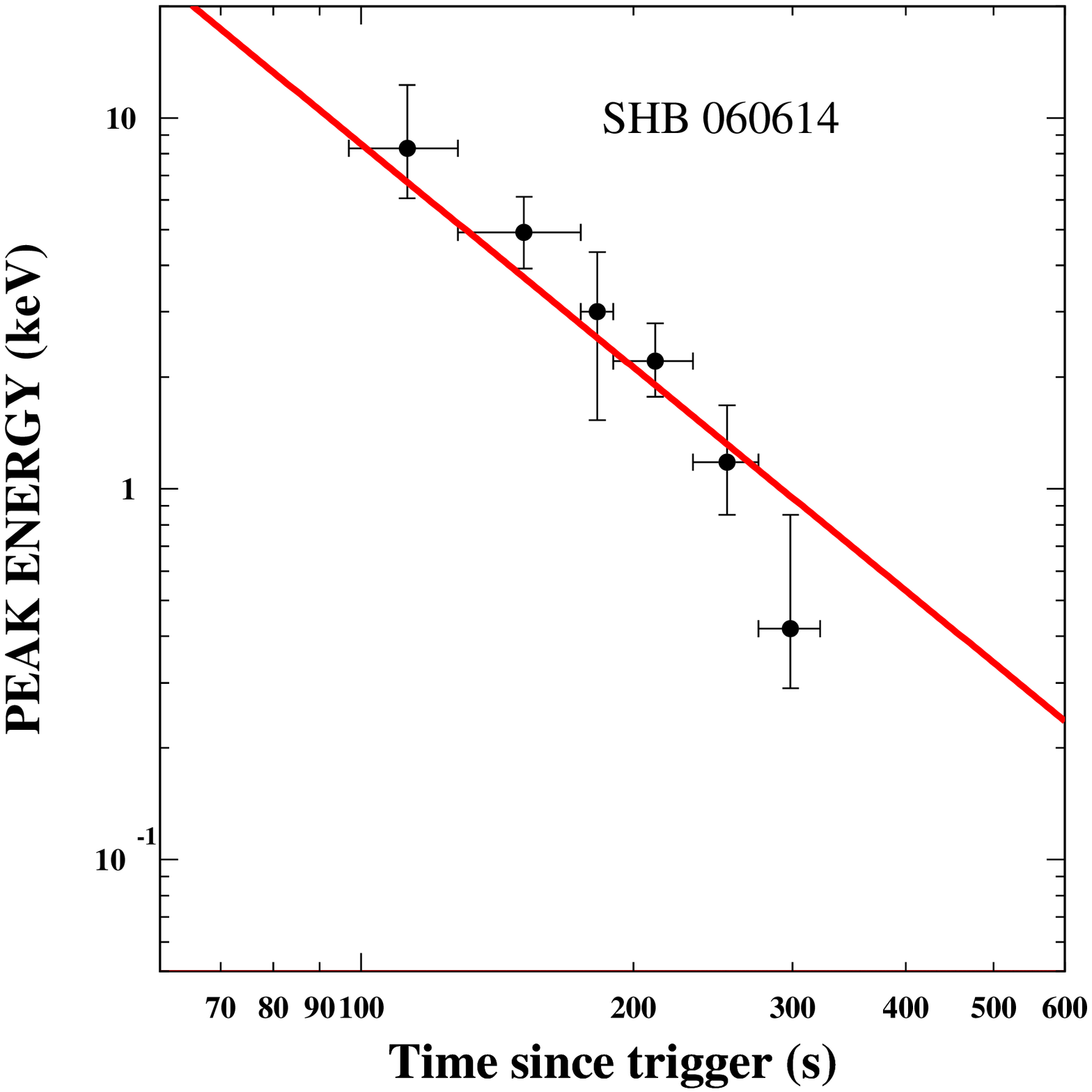,width=8.0cm,height=6.0cm }
}}
\caption{
Comparisons between observations 
and CB model predictions.
{\bf Top left (a):} The X-ray light curve of SHB 060614 (Evans et 
al.~2007).
{\bf Top right (b):} The R-Band light curve of SHB 060614 (Della Valle 
et al.~2006).
{\bf Middle left (c)}:The photon spectral index light curve of SHB 060614
(Zhang et al.~2007).
{\bf Middle right (d)}:The peak energy evolution during the fast
decay of the ESEC in SHB 061614
}
\label{F4}
\end{figure}

In Fig.~\ref{F4}a we present the CB model fit to the  canonical 
X-ray  light curve of GRB 060614. The decay of the ESEC is very well 
described by ICS of the light of a  superstar cluster
as the CB leaves the cluster and enters the ISM. 
Figs.~\ref{F4}b,c,d
compare the observed hardness  
ratio, evolution of the effective photon spectral index and the
evolution of $E_p$ during the fast decay phase  of the ESEC
and the CB model predictions. 
The agreement between theory and observations is quite satisfactory.
The XRT data on the decay of the prompt ICS emission 
extended only up to 480 s. The CB model estimates that it was taken over 
by SR only around 950 s. The second orbit XRT data began only at 
4.5 ks after the BAT trigger but clearly show 
a stretched plateau until 30 ks when it begins to bend into an asymptotic
power-law temporal decay.   
The plateau phase and the late decay of the X-ray afterglow 
are those expected for an SR afterglow from CB decelerating 
in an isothermal sphere  
density profile, $n\! \propto \!1/(r^2+r_s^2)$, namely
$F_\nu \! \sim\! t^{-(1+\beta_X)}\, \nu^{-\beta_X}$ with 
$\beta_X\!\approx \!1.1$
Such a density profile is  encountered by CBs which 
move from inside the host galaxy into its halo. 
The CB model best fit
value, $\gamma_0\, \theta\! = \!3.08$, is typical of XRFs, which, in the 
CB 
model, are interpreted as far off-axis GRBs (e.g., DD2004, DDD2004).
The corresponding value, $\delta\!=\!2\, \gamma/(1\!+\!\gamma^2\, 
\theta^2)\!\approx\!\gamma/5,$   yields $E_{iso}$ and $L_p$
which are, respectively,  $\!\sim \! 125$, and $625$
times smaller  
than when they are  measured at a  typical viewing angle, 
$\theta\!\approx\!1/\gamma_0$.

\section{Summary and main conclusions}
We have demonstrated that the entire observational data on SHBs can be 
explained by the assumption that SHBs are produced by highly relativistic 
jets ejected in processes involving compact stars, such as mergers of 
compact stellar objects, large-mass accretion episodes onto compact stars 
in close binaries or onto intermediate-mass black holes in dense stellar 
regions, and phase transition of compact stars. Natural environments of 
such events are the dense cores of globular clusters or superstar clusters 
and young supernova remnants. We have demonstrated that the cannonball 
model of GRBs can reproduce the main observed properties of their prompt 
emission, extended soft emission component and afterglows. In particular, 
we have used the CB model to fit the XRT light curve of all Swift SHBs 
with a well sampled X-ray afterglow. We have shown that their prompt 
gamma-ray emission is well described by inverse Compton scattering  
of the progenitor's glory light, their extended soft emission component by 
ICS of light of the host star cluster or by synchrotron radiation in the 
high density interstellar medium of superstar cluster, and their afterglow 
by synchrotron radiation outside the cluster. We have also demonstrated 
that nearby GRBs of low luminosity 
short end of the 
duration distribution of LGRBs and without an associated SN,  
such as GRBs 060614 and 060505, may be 
SHBs viewed far off axis. The BATSE observations on board CGRO indicated 
that nearly $\!sim\!$25\% of all GRBs are SHBs. The evidence from INTEGRAL 
that 
the rate of LGRBs is comparable to that of SNe of type Ib/c, which are a 
significant fraction of all core collapse SNe, implies that also the 
production rate of SHBs is large and comparable to the birth-rate of 
neutron stars. Hence neutron star merger, which has a rate much smaller 
than the birth-rate of neutron stars, cannot be the main source of SHBs. A 
high rate of SHBs (most of which are beamed away from us), comparable to 
that of SNe Ib/c, suggests that  phase transition in neutron stars or 
mass accretion episodes onto stellar and intermediate mass black holes are 
more probable sources of SHBs.

\noindent
{\bf Acknowledgment}: We would like to thank A. De R\'ujula
for a long and fruitful collaboration and  for his contributions  
to this manuscript. A.D. would like to thank the Theory Division 
of CERN for its hospitality during  this work.

\begin{deluxetable}{lllllc}
\tablewidth{0pt}
\tablecaption{Width and photon spectral index of SHBs and their ESEC}

\tablehead{
\colhead{SHB} & \colhead{$z$} & \colhead{$\Gamma$(SHB)}&
\colhead{Width[s]} & \colhead{${\rm T_{ESEC}[s]}$}&
\colhead{$\Gamma$(ESEC)}  }
\startdata

050509B & 0.225  &  $1.5\pm0.4$ &  0.03  &         &   \\                
050709  & 0.161  &  $0.8\pm~0.14$ &  0.22  &   131 & $1.98\pm 0.18$ \\
050724  & 0.258  &  $1.38\pm0.13$ &  0.256 &   152 & $2.5\pm 0.2$  \\       
050813  & 0.72   &  $1.2\pm 0.5$  &  0.6   &       &  \\                 
051105A &        & $1.33\pm 0.25$ &  0.028 &       &  \\
051210  & $>1.4$ & $1.0\pm 0.33$  &  1.27  &   40  &  \\               
051221A & 0.546  & $1.08\pm0.13$  &  1.4   &       &  \\                    
051227  &0.71 ?  & $1.09\pm 0.23$ &  0.6   &  110  &  \\                 
060121  & $>1.7$ & $0.82^{+0.38}_{-0.21}$&1.97&     &  \\
060313  & $<1.1$ & $0.61\pm 0.10$ &  0.7   &       &  \\
060502B & 0.287  & $1.0\pm 0.2$   & 0.09   &       &  \\
060505  & 0.089  & $1.3\pm 0.3$   &  4     &       &  \\
060614  & 0.125  & $1.57\pm 0.14$  &  5     &  178  & $2.13\pm 0.04$ \\
060801  & 1.131  & $1.3\pm 0.2 $   & 0.5    &       &   \\
061006  & 0.438  & $0.93\pm 0.07$  &  0.5   &  130  & $1.74 \pm 0.17$ \\
061201  & 0.111  & $0.81\pm 0.15$ & 0.8     &       &                 \\ 
061210  &0.41    & $0.79\pm 0.15 $ & 0.06    &   85  & $1.55\pm 0.28$ \\
061217  &0.827   & $0.96\pm0.28 $ & 0.22    &       &  \\
070209  &        & $1.02\pm 0.33 $ & 0.10    &       &  \\
070406  &        & $0.9 \pm 0.4 $   &   0.7   &       &  \\
070429B & 0.902  & $1.53 \pm 0.40$  &  0.5      &       &  \\
070714B & 0.922  & $0.86 \pm 0.10$  &  3    &   65    &$1.36\pm0.19$\\
070724A & 0.457  & $1.81 \pm 0.33$  &  0.40 &          &  \\
070729  &        & $1.08\pm 0.36$  & 0.9     &          &  \\
070809  & 0.2187 & $1.69\pm 0.22$  & 1.3     &            & \\
070810B &        & $1.44\pm 0.37$  & 0.08    &            &  \\
071112B &        & $0.69 \pm 0.34$  &  0.30   &            &  \\  
071227  & 0.383  & $0.99 \pm 0.22$  & 1.8     & 100        & \\
080123  &        & $1.22^{+0.21}_{-0.52}$  & 0.40  & 115  & 
$2.15~+\-~0.54$\\
\enddata

Based on data cited in Kann et al.~2008 and from
$http://gcn.gsfc.nasa.gov/gcn3_-archive.html\,. $

\label{t1}
\end{deluxetable}


\begin{deluxetable}{llllllllc}
\tablewidth{0pt}
\tablecaption{CB model parameters of SHBs' afterglows}
\tablehead{
\colhead{SHB} & \colhead{$t_1[s]$} & \colhead{$\Delta t_1[s]$}& 
\colhead{$t_b[s]$} & \colhead{$\theta\,\gamma_0 $}&
\colhead{$p^a$} & \colhead{$t_f[s]^b$} &
\colhead{$\Delta t_f[s]^b$} } 
\startdata
050709  & ---    & ---   & 46733& 0.716& (2.2) & ---  & ---  \\
050724  & 16.2   & 58.9  & 2170 & 0.327& 1.56 & 3138 &37244 \\
051210  &  0     & 92.3  & ---  & ---  & ---   & ---  & ---  \\
051221A &  0     & 63.9  & 11591& 1.15   & 2.2 & ---  & ---  \\
051227  &  85.0  & 24.0  & 1260 & 1.026  & 2.33 & --- & ---   \\
060121  &  ---   & ---   & 6754 & 1.031  & 2.2 & --- & ---  \\
060313  & 175.4  & 246.0 & 1442& 1.025 & 2.2 & 2.1 & 138   \\
060801  & 0      & 104.5 & --- & ---   &  ---  & --- & ---   \\
061006  & 0.17   &  89.9 & 8561& 0.778 & (2.2) & --- & ---   \\
070714B & 40.7   & 107.8 & 1672& 1.12  & (2.2) & 30.2& 49.2  \\
070724A & 87.8   & 108.9 & 1531& 1.483 & (2.2) & 58.9& 48.8  \\
071227  & 89.2   & 31.4  & 3093& 1.164 & (2.2) & 0   & 8.86  \\
080123  & 150.5  & 24.9  & 3501& 0.865 & (2.2) & 0   & 28.9  \\
080503  & 20.1   & 47.1  &218946&(1.00)& (2.2) & ---& ---    \\
060614  & 16.8   & 42.9  & 3829& 3.079 &  2.14 & --- & ---   \\
\enddata

\label{t2}
\end{deluxetable}

\newpage

\newpage

\newpage

\newpage

\end{document}